\documentclass[9pt,twocolumn,twoside]{pnas-new}
\usepackage{graphics,color,graphicx,amsmath}
\usepackage{xr}

\DeclareMathOperator{\sech}{sech}

\newcommand{\br}[1]{\left<#1\right>}

\newcommand{\te}{\tau_E}
\newcommand{\tildete}{\tilde\tau_E}
\newcommand{\pe}{p_E}
\newcommand{\tf}{{\tau_{\rm f}}}
\newcommand{\tc}{{\tau_{\rm c}}}
\newcommand{\he}{h_E}
\newcommand{\ta}{\tau_{\rm m}}
 
\templatetype{pnasresearcharticle} 

\title{Outsourcing Memory Through Niche Construction}

\significancestatement{All organisms must adapt to changing environments, but adaptation can modify the environment itself. We solve a version of this problem in terms of how long organisms remember. Shorter memory should be better for variable environments and longer for slow changing ones, but environmental variability depends on feedback. Surprisingly, we find the same mathematical law in both cases, revealing how much shorter memory should be relative to the environmental timescale. We consider how this depends on memory complexity and metabolic costs in populations, allowing us to predict a general set of conditions for when organism will outsource memory to the environment: when maintaining a brain is costly, the environment fluctuates quickly, and organisms inhabit a complex environment.}

\author[a]{Edward D. Lee}
\author[b]{Jessica C. Flack} 
\author[b]{David C. Krakauer}

\affil[a]{Complexity Science Hub Vienna, Josefst\ae dter Strasse 39, Vienna, Austria}
\affil[b]{Santa Fe Institute, 1399 Hyde Park Rd, Santa Fe, NM 87501}
\dates{This manuscript was compiled on \today}

\leadauthor{Lee} 

\authorcontributions{All authors helped develop the initial idea. E.D.L. did the analysis, modeling, and wrote the code. The authors drafted the manuscript jointly.}
\authordeclaration{The authors declare no competing interests.}
\correspondingauthor{\textsuperscript{2}To whom correspondence should be addressed. E-mail: edlee@csh.ac.at}


\begin{abstract}
Adaptation to changing environments is a universal feature of life and can involve the organism modifying itself in response to the environment as well as actively modifying the environment to control selection pressures. The latter case couples the organism to environment. Then, how quickly should the organism change in response to the environment? We formulate this question in terms of how memory duration scales with environmental rate of change when there are trade-offs in remembering vs.~forgetting. We derive a universal scaling law for optimal memory duration, taking into account memory precision as well as two components of environmental volatility, bias and stability. We find sublinear scaling with any amount of environmental volatility. We use a memory complexity measure to explore the strategic conditions (game dynamics) favoring actively reducing environmental volatility---outsourcing memory through niche construction---over investing in neural tissue. We predict stabilizing niche construction will evolve when neural tissue is costly, the environment is variable, and it is beneficial to be able to encode a rich repertoire of environmental states.
\end{abstract}

\begin{document}

\maketitle
\thispagestyle{firststyle}
\ifthenelse{\boolean{shortarticle}}{\ifthenelse{\boolean{singlecolumn}}{\abscontentformatted}{\abscontent}}{}

\dropcap{W}hat is the optimal timescale of adaptation---how long should memory of the environment persist when the environment is changing? And when should the organism invest in changing the rate of environmental change? Research in a wide range of fields suggests that bidirectional organism-environment feedback through niche construction and symbiosis is common and plays a significant role in shaping evolutionary dynamics. Slowly evolving genes co-evolve with quickly evolving culture \cite{feldman_gene-culture_1996}, as illustrated by the evolution of dairy-farming facilitating selection of alleles for adult lactase persistence \cite{gerbault_evolution_2011}. Quickly evolving organisms modify their otherwise slowly changing niches and alter selection pressures \cite{odling-smee_niche_1996, laland_cultural_2011, clark_niche_2020}, illustrated by yeast modifying fruit environments to attract Drosophilid flies that enhance yeast propagation \cite{buser_niche_2014}. Institutions feed back to influence individual decisions by changing cost landscapes and enhancing cultural transmission \cite{bowles_microeconomics_2006, poon_institutional_2022} (e.g.~legislation in support of same-sex marriage that increases the willingness to voice support in the face of risk \cite{ofosu_same-sex_2019}). To gain information about noisy, hidden variables and reduce social uncertainty, error-prone individual pigtailed macaques collectively compute a social power structure that reduces uncertainty about the cost of social interaction, making accessible new forms of conflict management (reviewed in references \citenum{flackCoarsegrainingDownward2017, ellis_lifes_2017}). Bacteria quorum sense, controlling group behavior in dynamically complex, changing environments (reviewed in reference \citenum{mukherjee_bacterial_2019}). Individuals, institutions, and firms all adapt to audit targets (Goodhart's Law), creating new feedbacks as they attempt to game the system \cite{merton_self-fulfilling_1948, strathern_improving_1997, soros_fallibility_2013, manheim_categorizing_2019}. In order to undermine competitors, agents can destabilize a system like in the recent Reddit-Gamestop event in which powerful hedge funds are thought to have introduced volatility to markets by manipulating Reddit users to short squeeze yet other hedge funds \cite{jakab_revolution_2022}. Motivated by these examples, we develop a synthetic framework that combines information theory, game dynamics, and scaling theory, in order to determine how adaptation scales in a range of plausible strategic settings including niche construction. 

We start by reformulating adaptation as rate of discounting of the past, building a conceptual and mathematical bridge to work on memory  \cite{kalmanNewApproach1960,brennerAdaptiveRescaling2000,gershmanStatisticalComputations2014,davisBiologyForgetting2017}. We take into account four factors: {\it bias} as preference in the environment for a particular state, {\it stability} as the rate at which environment fluctuates \cite{kussellPhenotypicDiversity2005,rivoireValueInformation2011}, {\it precision} as the capacity agents have to resolve environmental signal, and {\it feedback} as the rate of agent modification of the environment. In Table~\ref{tab:refs}, we provide examples of studies addressing the interaction of bias, stability, and precision. We also drop the separation of timescales assumption commonly made in modeling papers and explicitly consider feedback. We allow modification of the environment to be either passive or active, such that active modification can be destabilizing (increasing entropy) as well as stabilizing (reducing entropy). The Reddit-Gamestop event is one example of this ``destabilizing'' niche construction. Another is guerrilla warfare in which a weaker party randomly determines which battles to neglect by allocating zero resources \cite{chowdhury_experimental_2013}. In contrast, active agents can stabilize the environment by buffering against variation \cite{clark_niche_2020} or slowing its rate of change to reduce uncertainty about the future \cite{flackCoarsegrainingDownward2017, krakauer_information_2020}. A relatively simple example is stigmergy in which trails or routes are consolidated through repeated use \cite{theraulaz_brief_1999}. More complicated examples include the collective computation of slowly changing power structures in macaque groups \cite{brush_conflicts_2018} and foraging subgroup size distributions by spider monkeys \cite{ramos-fernandez_collective_2020} in which social structures are computed through communication and decision networks. Finally, we take into account how the precision \cite{ratcliffTheoryMemory1978} of an agent's or organism's estimates of environmental state influences its ability to fit the environment at a given degree of volatility.

\begin{table*}[t]\centering
\begin{tabular}{llll}
I Bias-Stability & II Stability-Precision & III Bias-Precision & IV Integrated\\
\hline
Taxis of larval invertebrates \cite{koehlSwimmingUnsteady2015} & Seed dormancy/germ banking    \cite{evansGermBanking2005} & Bandit problems    \cite{slivkinsIntroductionMultiarmed2019} & Volatile bandits    \cite{kaspiLevyBandits1995}\\
Stochastic voting models \cite{schofieldDivergenceSpatial2008} & Particle swarms \cite{yuanParticleSwarm2010} & Microbial chemotaxis      \cite{tindallTheoreticalInsights2012} & Learning changing data sources   \cite{bifetLearningTimeChanging2007}\\
Learning changing distributions  \cite{kosinaHandlingTime2012} & Cognitive aging  \cite{rollsStochasticCortical2015} & Speed-accuracy trade-offs   \cite{ratcliffModelingResponse1998,bruntonRatsHumans2013,mileticSpeedAccuracy2019} & Consensus with link failure \cite{karDistributedConsensus2009}\\
Loss/Change aversion      \cite{schweitzerDecisionBias2000} &  & Optimal foraging   \cite{tregenzaBuildingIdeal1995} & Retinal sensitivity rescaling \cite{brennerAdaptiveRescaling2000} \\
  &  & Page Rank consensus     \cite{musaComparingRanking2018} & \\
\end{tabular}
\vspace{.05in}
	\caption{Classification of studies in terms of bias, stability, and precision. We group studies according to the pairs of factors they combine (I) bias-stability, (II) bias-precision, and (III) stability-precision. Studies that implicitly combine all of these factors are noted under ``integrated'' (IV). Studies in category I focus on rules that apply in variable environments (bias) where environmental distributions are prone to rapid change (stability). Studies in category II focus on rules that apply when environments are likely to change (stability) and where making the correct decision depends on sensitivity to signals (precision or also ``accuracy'' in some literature). Studies in category III focus on rules that apply in variable environments (bias) and where making the correct decision depends on power of sensors to detect signal (precision). Studies in category IV include elements of I--III and apply to variable environments prone to rapid change, where sensory precision varies. Feedback with the environment, where agent inference modifies the input statistics and timescales are formally coupled, remains little considered despite being a central premise of research on niche construction and stigmergy. }\label{tab:refs}
\end{table*}

In ``Result 1,'' we explore the conditions under which long memory is beneficial. In ``Result 2,'' we derive the scaling relationship for optimal memory duration and environmental change. In ``Result 3,'' we derive by way of a back-of-the-envelope calculation the costs of memory using the literature on metabolic scaling. In ``Result 4,'' we introduce game dynamics introducing a complexity cost of memory to explore the evolution of active modification and outsourcing of memory to the environment.

\section*{Model structure \& assumptions}
\label{model}
\begin{figure}[tb]\centering
	\includegraphics[width=\linewidth]{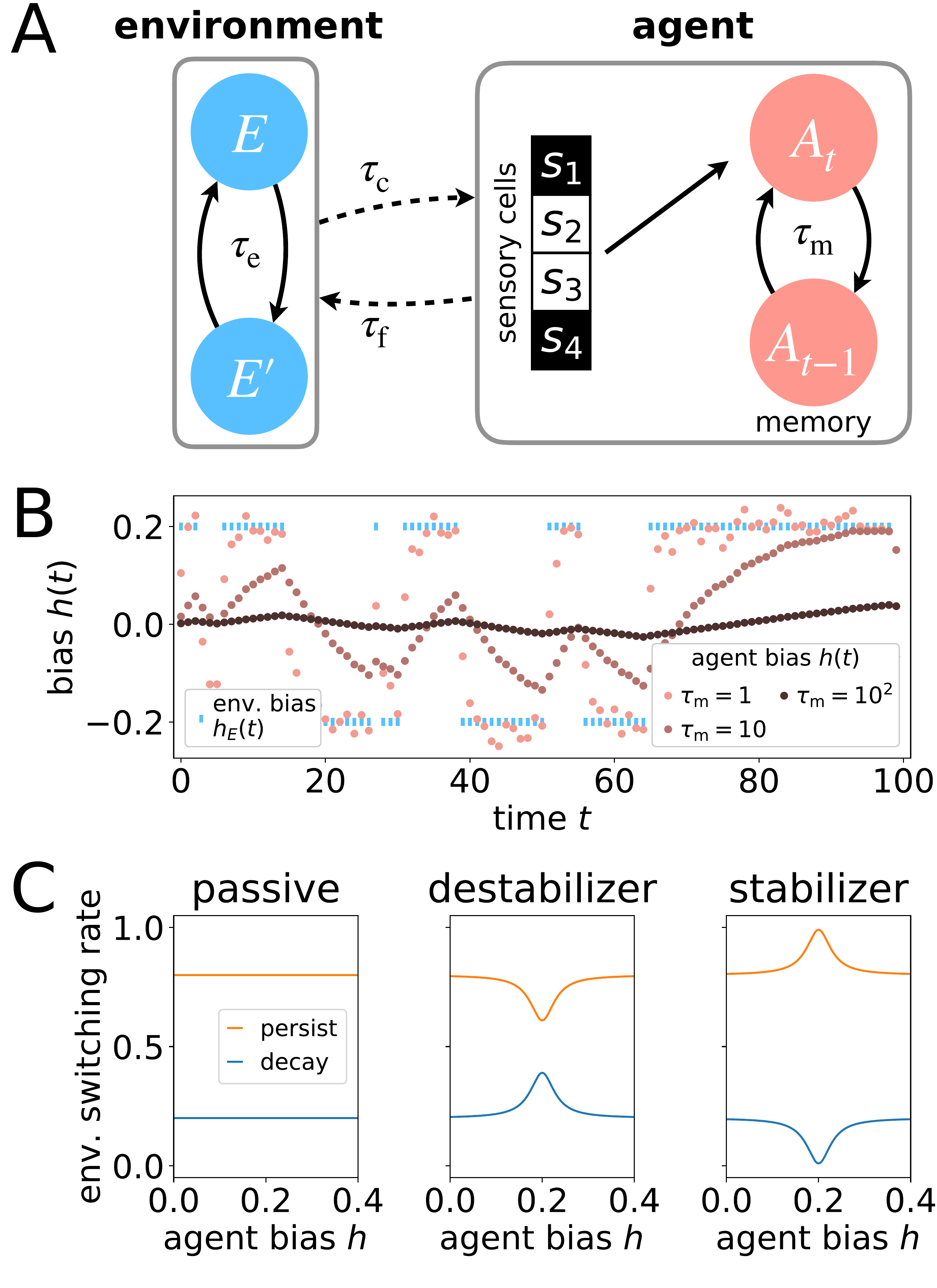}
	\caption{(A) Overview of framework. Environment $E$ switches configuration on timescale $\te$. The agent measures the current environment through sensory cells with precision $\tc$, here worth 4 bits. To obtain an estimate of environment statistics at time $t$, the agent $A_t$ combines present sensory estimates with memory of previous estimates recorded in an aggregator $A_{t-1}$ (Eq~\ref{eq:memory}) such that memory decays over time $\tau_{\rm m}$ (Eq~\ref{eq:agent memory}). Coupling with the environment speeds up or slows down environmental change on timescale $\tf$ (Eq~\ref{eq:rate}). (B) Example trajectories of agents adapting to environmental state $\he(t)$ with short, medium, and long memory. (C) Rate of environment switching per time step as a function of agent bias $h$ relative to environmental bias $\he=0.2$. For passive agents, switching rate does not depend on agent bias. For destabilizers $\alpha=0.95$, for stabilizers $\alpha=-0.95$. For both, $v=0.1$ from Eq~\ref{eq:rate} and environmental timescale $\te=5$.}\label{gr:overview}
\end{figure}

We summarize the structure of our model in Figure~\ref{gr:overview}, which combines the essential components of adaptive agents. As a result, it connects passive agents that learn the statistics of a fluctuating environment with those that modify the environment itself. We summarize notation in Appendix Table~\ref{si tab:notation}.

The environment $E$ at time $t$ is described by a probability distribution $\pe(s,t)$ over configurations $s$, a vector of descriptive properties. The environment has a bias for preferred states that changes on a relatively slow timescale. Here, we represent the state of the environment with a single bit $s\in\{-1,1\}$, analogous to the location of a resource as a choice between left and right \cite{couzinEffectiveLeadership2005,franksReconnaissanceLatent2007,bogaczPhysicsOptimal2006,bruntonRatsHumans2013}. In one configuration, the distribution of resources $\pe$ is biased to the left at a given time $t$, or $\pe(s=-1,t)>\pe(s=1,t)$, such that an agent with matching preference would do better on average than an agent with misaligned preference. In the mirrored configuration, the environment shows a bias of equal magnitude to the right $\pe(s=-1,t)<\pe(s=1,t)$. Such probabilistic bias can be represented as an evolving ``field'' $\he(t)$,
\begin{align}
	\pe(s,t) &= \frac{1}{2} + \frac{s}{2}\tanh{\he(t)},\label{eq:field}
\end{align}
such that reversal in bias corresponds to flip of sign $\he(t)\rightarrow-\he(t)$ that naturally embodies a symmetry between left and right. At every time point, the environment has clearly defined bias in one direction or another, determined by setting the external field to either $\he(t)=-h_0$ or $\he(t)=h_0$. With probability $1/\te$ per unit time, the bias in the environment reverses such that over time $\te$ the environment remains correlated with its past. When $\te$ is large, we have long correlation times and a slow environment or a ``slow variable.'' This formulation yields a stochastic environment whose uncertainty depends on both fluctuation rate, such that low rate implies high stability, and the strength of bias for a particular state, such that a strong bias yields a clear environmental signal. 

Passive agents sample from the environment and choose a binary action. In principle, the precision of the choice is dependent on the number of sensory cells contributing to the estimate of environmental state, the sensitivity of those cells, and the number of samples each cell collects while the contribution of each factor to the estimate can differ. In our model, all the alternatives are captured by $\tc$. When $\tc$ is high (either because the sensory cells sampled from the environment for a long time, many sensory cells contributed estimates, or each sensory cell is very sensitive) agents obtain exact measurements of the environment. A small $\tc$ corresponds to noisy estimates. The resulting estimate of environmental state $\hat p$ thus incurs an error $\epsilon_\tc$, 

\begin{align}
	\hat p(s,t) &= \pe(s,t) + \epsilon_\tc(t).\label{eq:bias}
\end{align}
From this noisy signal, sensory cells obtain an estimate of bias $\hat h(t)$, which is related to environmental bias $\he(t)$ plus measurement noise $\eta_\tc(t)$,
\begin{align}
	\hat h(t) &= \he(t) + \eta_{\tc}(t).\label{eq:h(t)}
\end{align}
In the limit of large precision $\tc$ and given that the noise in the estimated probabilities $\epsilon_\tc(t)$ from Eq~\ref{eq:bias} is binomial distributed, the corresponding error in field $\eta_\tc(t)$ converges to a Gaussian distribution (see Materials and Methods). Then, at each time step the agent's measurement of the environment includes finite-sample noise which is inversely related to precision.
 
An aggregation algorithm determines how much to prioritize the current measurement over historical ones. This gives the duration of memory by recording the agent's estimate of the state of the environment at the current moment in time $h(t)$ and feeding it to sensory cells at time $t+1$ with some linear weighting $0\leq \beta\leq 1$ \cite{mcnamaraMemoryEfficient1987},
\begin{align}
	h(t+1) &= (1-\beta) \hat h(t+1) + \beta h(t).\label{eq:memory}
\end{align}
This estimate is stored in an ``aggregator'' $A_{t}$, and we define $h(0)=0$. The weight $\beta$ determines how quickly the previous state of the system is forgotten such that when $\beta=0$ the agent is constantly learning the new input and has no memory and when $\beta=1$ the agent ceases to learn preserving its initial state. In between, agent memory decays exponentially with lifetime
\begin{align}
	\ta\equiv -1/\log\beta. \label{eq:agent memory}
\end{align}
We think of the weight $\beta$ that the aggregation algorithm places on the current estimate relative to the stored value as the timescale of adaptation $\tau_{\rm m}$, or agent memory duration. 

The output of this computation is the agent's behavior, $p(s,t)$. We measure the effectiveness of adaptation, or fit to the environment, with the divergence between a probability vector describing an agent and that of the environment. Measures of divergence, like Kullback-Leibler (KL) divergence, and, more generally, mutual information, have been shown to be natural measures of goodness of fit in evolutionary and learning dynamics from reinforcement learning through to Bayesian inference \cite{donaldson-matasci_fitness_2010, krakauer_darwinian_2011}.

\begin{figure*}[tb]\centering
	\includegraphics[width=.7\linewidth]{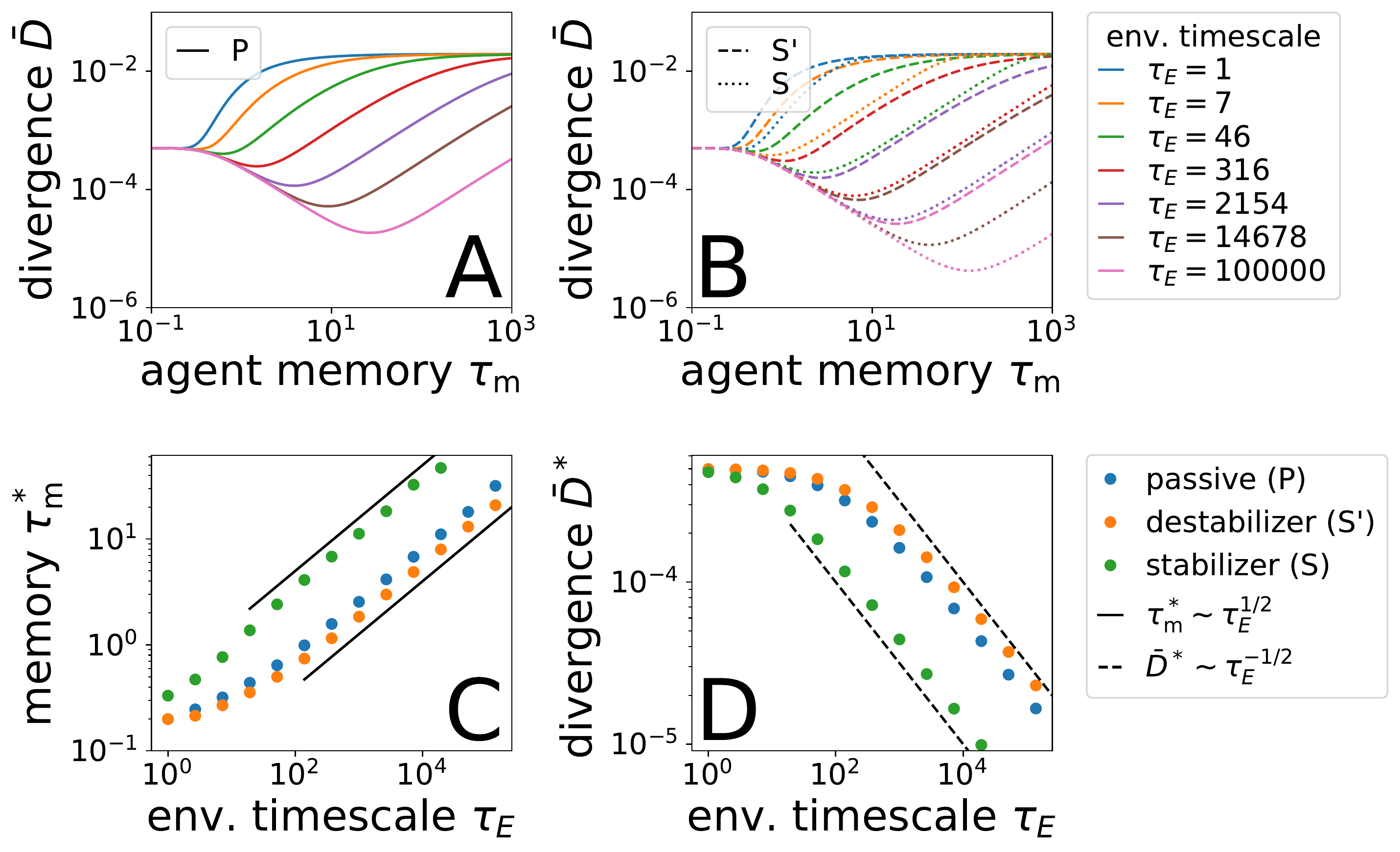}
	\caption{Divergence $\bar D$ as a function of agent memory $\ta$ and environmental timescale $\te$ for (A) passive and (B) active agents including destabilizers $S'$ and stabilizers $S$. For longer $\te$, agents with longer memory always do better, a pattern emphasized for stabilizers and diminished for destabilizers. (C) Scaling of optimal memory duration $\ta^*$ with environmental timescale $\te$, corresponding to minima from panels A and B. (D) Divergence at optimal memory duration $\bar D^*\equiv \bar D(\ta^*)$. Environmental bias $h_0=0.2$.}\label{gr:tau scaling}
\end{figure*}

Here we extend the model to include feedback by allowing agents to alter environmental stability, which is operationalized as the probability of switching. We add to the switching rate $1/\te$, the {\it active} construction rate,
\begin{align}
	\frac{1}{\tf(t)} \equiv \frac{v^2/\te}{[h(t)-\he(t)]^2 + v^2}, \label{eq:rate}\\
\intertext{such that the probability $q$ that the environment changes at the next point in time is}
	q[\he(t+1)\neq\he(t)] = 1/\te + \alpha/\tf(t). \label{eq:prob of change}
\end{align}
Eq~\ref{eq:rate} is written so that it remains normalized for arbitrary $v$ and that the rate gets smaller as the squared distance between agent bias and environmental bias $[h(t)-\he(t)]^2$ goes to zero. The probability $q$ of the environment switching to the opposite configuration includes weight $\alpha \in (0,1]$ to tune the strength of destabilizers, or $\alpha \in [-1,0)$ for stabilizers. This means that for positive $\alpha$, the rate of switching increases as the agent matches the environment more closely and the opposite for negative $\alpha$, whereas the parameter $v$ controls how closely the agent must match the environment to have an effect (i.e.~the width of the peak as plotted in Figure~\ref{gr:overview}C). The two types of active agents capture two ways adaptive behavior can feedforward to influence the timescale environmental change.\footnote{Note that in this binary example, the new environmental configuration when switching is unique, enforcing a deterministic switch, but in general there may be a large number of $K\gg1$ options such that the agent cannot easily guess at the results of environmental fluctuations.} We note that when $1/\tf=0$, we obtain {\it passive} agents that do not modify their environment, thus connecting passive and active agents to one another along a continuum scale.

Putting these elements of adaptation together, as shown in Figure~\ref{gr:overview}A, we obtain a toy learning agent that infers the statistics of a time-varying and stochastic environment.

\section*{Result 1: Long memory and adaptation favored when sensory cells are imprecise \& environments are slow}
The timescale of adaptation represents a balance between the trade-offs of preserving an internal state for too long or losing it too fast. We explore this trade-off by calculating an agent's fit to a changing environment. The fit can be quantified with the KL divergence between environment $\pe(s,t)$ with bias $h_{E}(t)$ and agent $p(s,t)$,
\begin{align}
	D_{\rm KL}[\pe||p](t) &= \sum_{s\in\{-1,1\}} \pe(s, t)\log_2 \left(\frac{\pe(s,t)}{p(s,t)}\right). \label{eq:DKL}
\end{align}
When the KL divergence is $D_{\rm KL}=0$, the agents use optimal bet-hedging, known as ``proportional betting,'' which is important for population growth dynamics \cite{kellyNewInterpretation1956,coverElementsInformation2006}. Eq~\ref{eq:DKL} is also minimized for Bayesian learners under optimal encoding \cite{contiNonequilibriumDynamics2020}. Assuming agents are playing a set of games in which they must guess the state of the environment at each time step, Eq~\ref{eq:DKL} is the information penalty paid by imperfect compared to perfect agents. 

After averaging over many environmental bias switches, we obtain the agent's typical {\it divergence},
\begin{align}
	\bar{D} &\equiv \lim_{T\rightarrow\infty} \frac{1}{T} \sum_{t=0}^{T-1} D_{\rm KL}[\pe||p](t),\label{eq:D}
\end{align}
The bar notation signals an average over time. Thus, fit improves as $\bar D$ decreases. 

In Figures~\ref{gr:tau scaling}A and B, we show divergence $\bar D(\ta,\te)$ as a function of the agent's memory $\ta$ given environmental timescale $\te$. In the limiting cases in which an agent has either no memory and is constantly adapting or has infinite memory and adaptation is absent, the timescale on which environmental bias changes ultimately has no effect---we observe convergence across all degrees of bias and stability. When an agent has no memory, or $\ta=0$, an agent's ability to match the environment is solely determined by its sensory cells. Low precision $\tc$ leads to large errors on measured environmental bias $\he(t)$ and large divergence $\bar D(\ta=0)$. On the other hand, high precision $\tc$ increases performance and depresses the intercept (Eq~\ref{eq:D at 0}). At the right hand side of Figure~\ref{gr:tau scaling}A, for large $\ta\gg1$, behavior does not budge from its initial state. Assuming that we start with an unbiased agent such that the transition probability is centered as $q(h) = \delta(h)$, the Dirac delta function, the agent's field is forever fixed at $h=0$. Then, divergence $\bar D(\ta=\infty)$ reduces to a fixed value that only depends on environmental bias (Eq~\ref{eq:divergence  at infty}). In between the two limits of zero and infinite agent memory, the model produces a minimum divergence $\bar D(\ta=\ta^*)$. This indicates the optimal duration of memory $\ta^*$ for a given degree of environmental bias and stability.

\begin{figure}[tb]\centering
	\includegraphics[width=.9\linewidth]{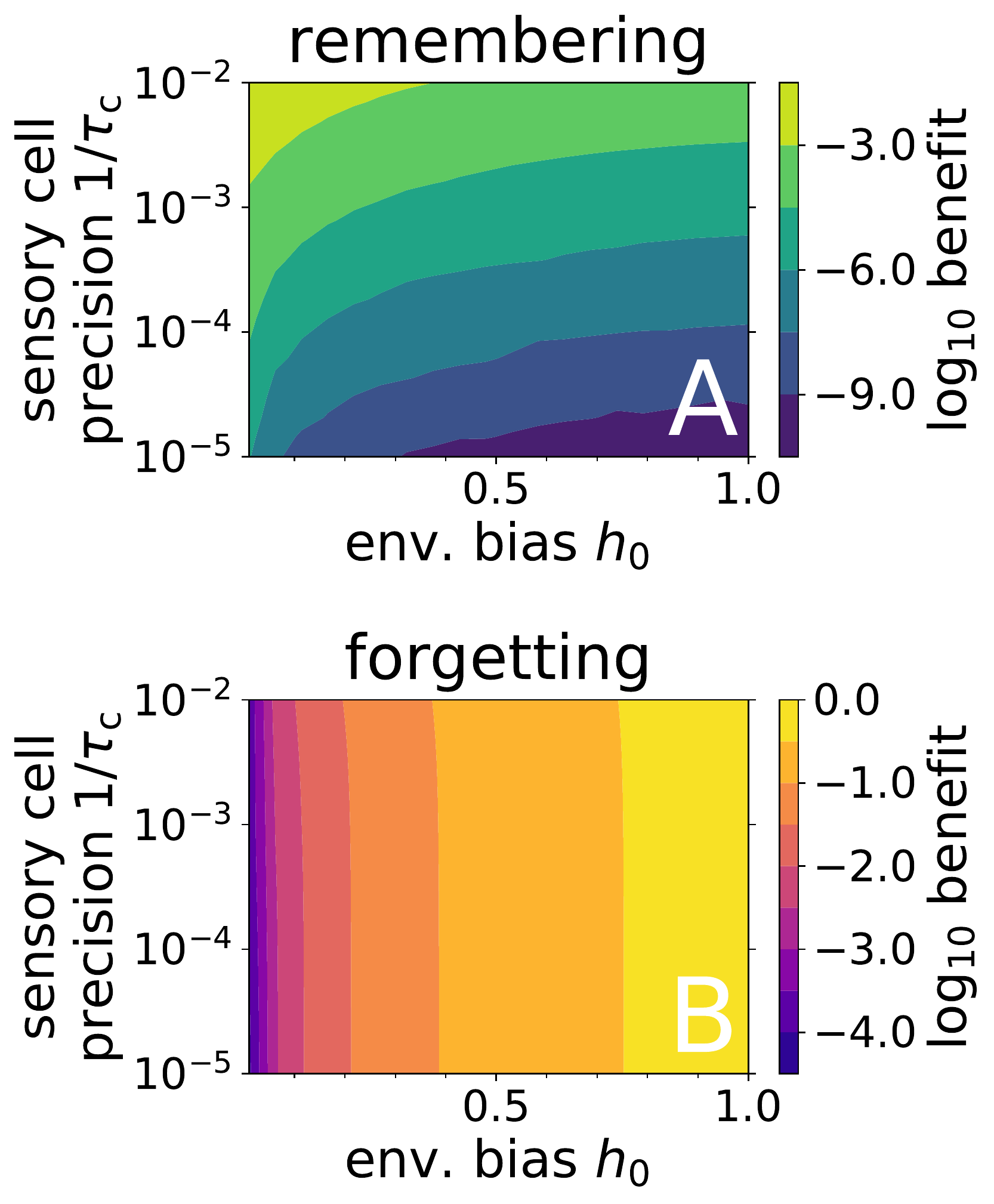} 
	\caption{Benefit from (A) remembering and from (B) forgetting defined as the reduction in divergence at optimal memory duration relative to no memory, $\bar D(\ta=0)-\bar D(\ta^*)$, and optimal memory duration to infinite memory, $\bar D(\ta=\infty)-\bar D(\ta^*)$, respectively. We show passive agents given environmental timescale $\te=10$. All contours must converge (A) when $h_0=0$ and (B) when $\tc=0$. Agent-based simulation parameters specified in accompanying code.}\label{gr:remembering v forgetting}
\end{figure}

The benefits of memory are more substantial for agents with imprecise sensory cells. This benefit is the difference $\bar D(\ta=0)-\bar D(\ta=\ta^*)$ as shown in Figure~\ref{gr:remembering v forgetting}A. As one might expect, integrating over longer periods of time provides more of a benefit when the present estimate $\hat p$ is noisy, $\tc^{-1}$ is large, and sensory cells are not particularly precise, a deficiency in precision that memory counters by allowing organisms to accumulate information over time. This intuition, however, only applies in the limit of large environmental bias $h_0$ where the contours of optimal memory flatten and become orthogonal to precision $\tc^{-1}$. When the bias in the environment is weak, the curved contours show that the benefits of memory come to depend strongly on nontrivial interaction of precision and environmental bias. The complementary plot is the benefit from forgetting, $\bar D(\ta=\infty)-\bar D(\ta=\ta^*)$ in Figure~\ref{gr:remembering v forgetting} B, which is largely determined by bias $h_0$. When bias is strong, the costs of estimating the environment inaccurately are large, and it becomes important to forget if sensory cells are imprecise. Thus, our model encapsulates the trade-off between remembering and forgetting both in terms of their absolute benefits as well as the emergence of simple dependence of the respective benefits in the limits of high environmental bias and high sensory precision. An agent has optimally tuned its timescale of adaptation $\ta=\ta^*$ when it has balanced the implicit costs of tuning to fluctuations against the benefits of fitting bias correctly.

\section*{Result 2: Adaptation and environmental change scale sublinearly}
For sufficiently slow environments, or sufficiently large $\te$, we find that optimal memory duration $\ta^*$ scales with the environmental timescale $\te$ sublinearly as in Figure~\ref{gr:tau scaling}C. To derive the scaling between optimal memory and environmental timescale, we consider the limit when agent memory persistence is small relative to the environmental persistence $\ta\ll\te$. Under this condition, optimal memory represents a trade-off between a poor fit lasting time  $\ta$ and a good fit for time $\te-\ta$. During the poor fit, the agent pays a typical cost at every single time step such that the cost grows linearly with its duration, $\mathcal{C} \ta$, for constant $\mathcal{C}$. When the environment is stable, agent precision is enhanced by a factor of $\ta$ because it effectively averages over many random samples, or a gain of $\mathcal{G} \log \ta$ for constant $\mathcal{G}$. When we weight each term by the fraction of time spent in either transient or stable phases, $\ta/\te$ and $(\te-\ta)/\te$ respectively, we obtain the trade-off
\begin{align}
	\mathcal{C}\frac{\ta^2}{\te} - \mathcal{G} \frac{\te-\ta}{\te} \log\ta.\label{eq:tau scaling trade-off}
\end{align}
At optimal memory $\ta^*$, Eq~\ref{eq:tau scaling trade-off} will have zero derivative. Keeping only the dominant terms and balancing the resulting equation, we find
\begin{align}
	\ta^* &\sim \te^{1/2}. \label{eq:tau*}
\end{align}
This scaling argument aligns with numerical calculation as shown in Figure~\ref{gr:tau scaling}C. 

Similarly, we calculate how optimal divergence  $\bar{D}^*$ scales with environmental timescale. Assuming that the agent has a good estimate of the environment such that the error in average configuration $\epsilon_\tc(t)$ is small, agent behavior is $\pe(s,t) + \epsilon_\tc(t)$ and $\epsilon_\tc(t)$ is normally distributed. Then, we expand the divergence about $\pe(s,t)$ in Taylor series of error $\epsilon_\tc(t)$ (Materials \& Methods). Over a timescale of $\ta^*$, the precision of this estimate is further narrowed by a factor of $\ta^*$ such that
\begin{align}
	\bar D^* \sim 1/\ta^* \sim \te^{-1/2}.\label{eq:D scaling}
\end{align}
Although we do not account for the transient phase, we expect the relation in Eq~\ref{eq:D scaling} to dominate in the limit of large $\te$, and our numerical calculations indeed approach the predicted scaling in Figure~\ref{gr:tau scaling}C.\ In contrast, when environment does not fluctuate, or bias $h_0=0$, agents pay no cost for failing to adapt to new environments and infinite memory is optimal. Overall, the sublinear scaling between memory duration and rate of environmental change indicates an economy of scale. Agents require proportionally less expenditure on adaptation in slow environments than would be true under a linear relationship. Hence a slow environment is in this sense highly favorable to an adaptive agent when considering the costs of poor adaptation. 

\section*{Result 3: Metabolic cost of memory can become prohibitive in slow environments}
Here we ask how memory might become limited by the metabolic costs of  neural tissue. 

\begin{figure}[tb]\centering
	\includegraphics[width=.9\linewidth]{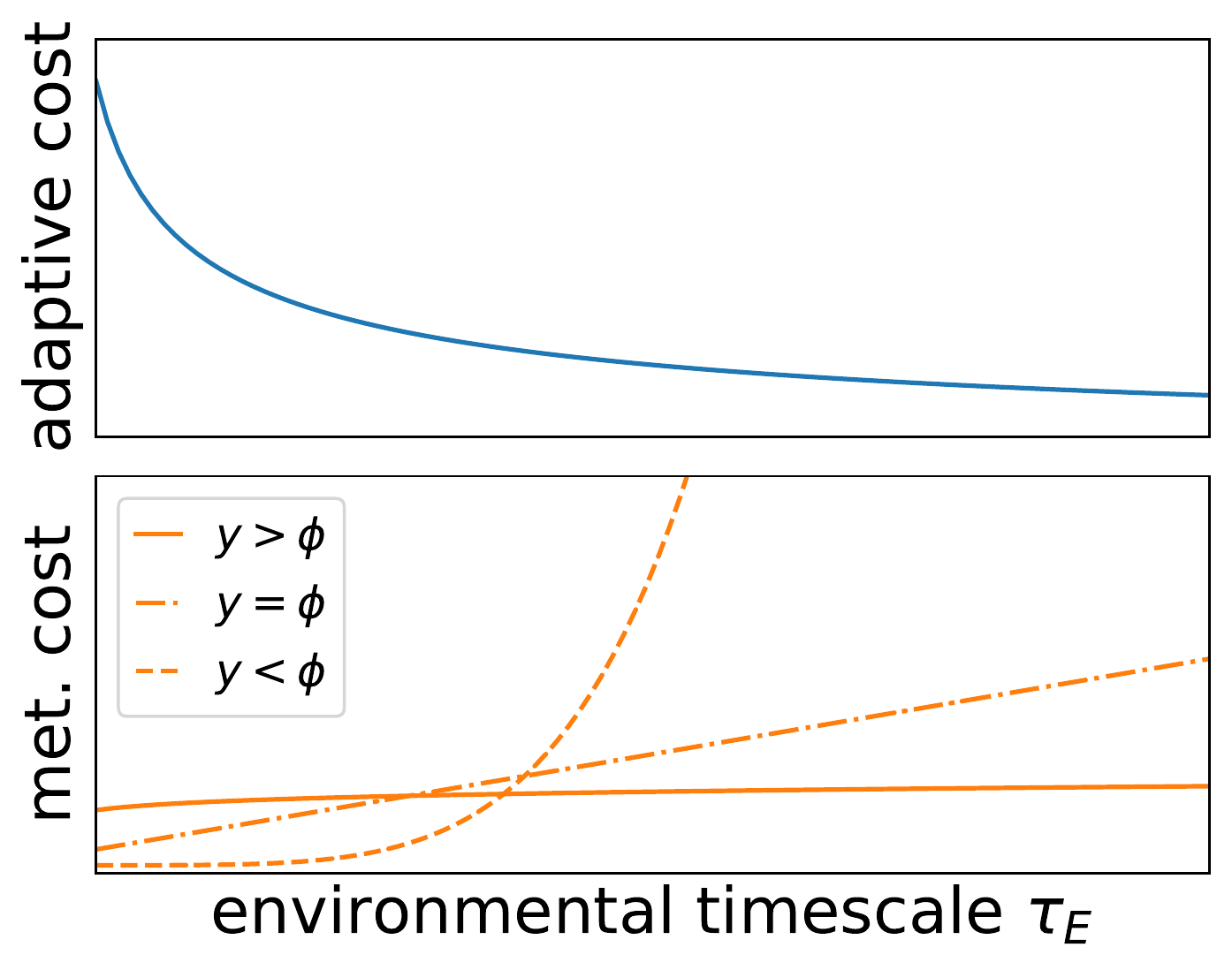}
	\caption{Scaling of adaptive and metabolic costs with environmental timescale $\te$. (A) Adaptive cost $\bar{D}$ is largest at small $\te$, but (B) metabolic costs are largest for longer-lived organisms with scaling dependent on exponents $y$ and $\phi$ such as for ``elephants'' that experience slower environments (Eq~\ref{eq:metabolic scaling}).}\label{gr:metabolism}
\end{figure}

We start with the well-documented observation that physical constraints on circulatory networks responsible for energy distribution influence organismal traits including lifespan and size across the animal kingdom from microorganisms to blue whales \cite{westFourthDimension1999,whiteMammalianBasal2003}. Metabolic costs matter for brain mass $M_{\rm br}$, which scales with body mass $M_{\rm bo}$ sublinearly, $M_{\rm br} = A M_{\rm bo}^a$, where $a=3/4$ across taxa (within individual taxa it spans the range 0.24 to 0.81 \cite{burgerAllometryBrain2019}). To account for memory cost, we make the simple assumption that the quantity of brain mass required for memory is proportional to the number and duration of environmental states (the ``environmental burden'') the organism encounters,
\begin{align}
	M_{\rm br} &\propto N \te. \label{eq:allo2}
\end{align}
After all, we say, ``an elephant never forgets'' and not the same of a mouse. 

Now, we use predictions of allometric scaling theory to relate metabolic rate $B$ to mass, $B \propto M^{1/4}$ \cite{westAllometricScaling2002}, and lifespan to body mass, $T \propto M_{\rm bo}^{b}$ for metabolic exponent $b=1/3$ \cite{savageSizingAllometric2008}.  From Eq~\ref{eq:allo2}, we obtain a relationship between metabolic rate and memory burden, $B \propto N^{\phi}\te^{\phi}$, where $\phi \equiv a/4b$.\footnote{When we use $a=3/4$, we obtain the range $\phi = [5/8,15/16]$, the endpoints depending on whether $b=0.3$ or $b=0.2$, respectively, while accounting for taxa-specific variation in $a$ leads to much wider range of $\phi \in [0.2,1.01]$. Thus, we hypothesize that longer environmental timescales lead to increased brain mass and metabolic expenditure with sublinear scaling.} Note that this scaling is sublinear for biological organisms, $\phi<1$. Although the adaptive cost decays with $\te$ in Figure~\ref{gr:metabolism}A, metabolism grows as $\te^\phi$ as shown in Figure~\ref{gr:metabolism}B. The competing scalings suggest that for small organisms the cost of adaptation will make a disproportionate contribution to the lifetime energy budget of an organism. This is consistent with observations on developmental neural growth in butterflies \cite{snell-roodBrainSize2009}.\footnote{As noted in the cited study and its citations, experience leads to larger brain size, indicating that learning from such experience is sufficiently valuable to warrant concomitant constitutive and induced costs.}

To generalize the previous argument, we assume larger organisms experience longer environmental timescales. Then, $\te\propto T^y$, where $y\in[0,1]$ to ensure that $\te$ and $N$ increase together since $\te^{1/y-1}\propto N$. We now find the relationship between metabolic rate and environmental timescale
\begin{align}
	B \propto \te^{\phi/y} \propto N^{\phi/(1-y)},\label{eq:metabolic scaling}
\end{align}
which reduces to the previous case when $y=1$ (and $N$ is a constant). Such dependence implies that the metabolic cost of memory will explode with environmental timescale (and organism lifetime) as $y$ approaches zero and grow slowly and sublinearly when $y=1$.  Both possibilities are shown in Figure~\ref{gr:metabolism}B. More generally, lifespan is expected to influence the relative contributions of adaptive versus metabolic costs  \cite{lieftingWhatAre2019,woudeNoGains2019}.

\section*{Result 4: Niche construction, memory complexity, \& the outsourcing principle}
\label{game}
In Result 3, we explored the metabolic cost of memory versus adaptation, emphasizing the metabolic constraints on long memories. In this section we focus on the information costs of adaptation when allowing for active modification of the environment.  We explore how outsourcing memory to the environment by slowing it down is beneficial when costs of poor adaptation are dominant \cite{flackCoarsegrainingDownward2017}. 

A slow environmental timescale increases the advantages of persistent memory, but it also reduces the amount of new information an organism requires by reducing uncertainty about the state of the environment. 
In this sense, slow environmental variables reflect a form of niche construction. Whether ant pheromone trails, food caching, collectively computing power structures, writing, or map-making, niche construction that promotes the stability or predictability of the local environment \cite{clark_niche_2020,klyubinTrackingInformation2004} reduces the number of environmental configurations that an organism needs to encode. 
Stabilizing niche construction, however, also creates a public good that by reducing environmental uncertainty, provides a benefit to all agents, and can be exploited by free riders. This can lead to a tragedy of the commons \cite{krakauerDiversityDilemmas2009}.

\begin{figure}\centering
	\includegraphics[width=\linewidth]{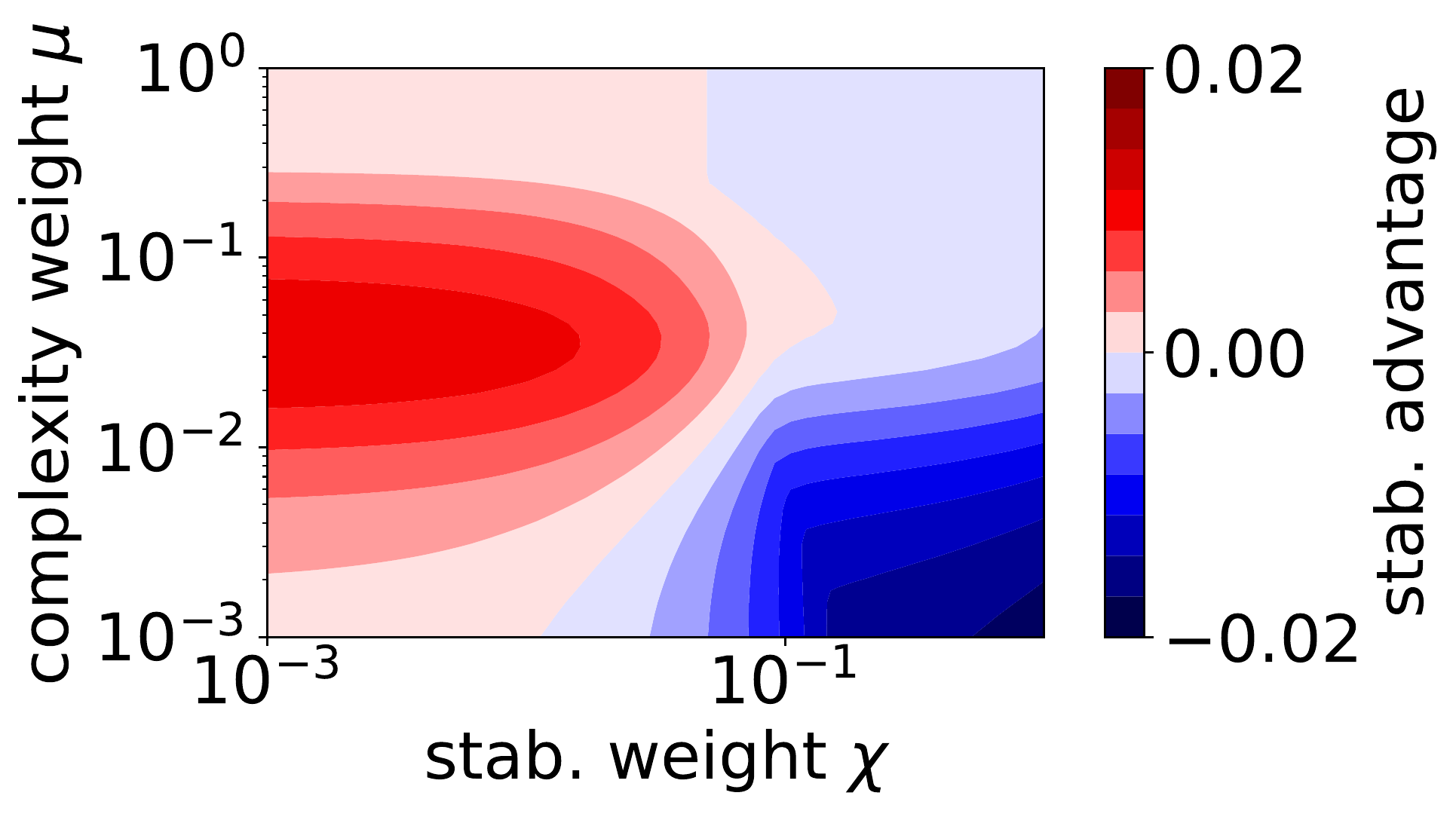}
	\caption{Comparison of total divergence for stabilizers $\mathcal{D}_{S}$ and destabilizers $\mathcal{D}_{S'}$, or $\mathcal{D}_{S'} - \mathcal{D}_{S}$, in a fixed environment and common sensory precision and costs. The difference is between agents poised at optimal memory duration given $\mu$, $\chi$, and $\beta$. Small stabilization weight $\chi$ favors stabilizers, whereas high monopolization cost $\mu$ favors destabilizers. Simulation parameters are specified in accompanying code. }\label{gr:fitness}
\end{figure}

We explore the conditions under which active modification of the environment can evolve given the free-rider problem, and how this overcomes the costs of adaptation. We introduce stabilizing mutants into a population of passive agents. Assuming other organisms are poorly adapted to regularities in the environment, we expect stabilizing mutants to gain a competitive advantage but only over the short term. In the longer term, an established stabilizer population is susceptible to invasion by free-riders exploiting outsourced memory; said another way, stabilizers slow environmental timescales and reduce divergence for all individuals sharing the environment, but they uniquely pay for stabilization. Thus, as in the classical example of niche construction, the usual ``tragedy of the commons'' argument makes it an evolutionary dead end \cite{krakauerDiversityDilemmas2009}. 

It follows that stabilization is only a competitive strategy if individuals can monopolize extraction of resources from the stabilized environment. In the natural world, this could occur through physical encryption (e.g.~undetectable pheromones \cite{wenBreakingCipher2017}), the erasure of signal (e.g.~food caching \cite{smithEvolutionFood1984}), or the restriction of social information (e.g.~concealment \cite{hallChimpanzeeUses2017}).
 To model competition between monopolistic stabilizers and other strategies, we account for the costs of memory, stabilization, and precision. We introduce a new memory cost of encoding complex environments as
\begin{align}
	H(\ta) &= \log_2 (1+1/\ta).\label{eq:memory cost}
\end{align}
Eq~\ref{eq:memory cost} can be thought of as a cost of exploring more configurations over a short period time versus agents that are temporally confined. This is different from costs associated with the environmental burden in Result 3, which emphasizes the costs of persistence, not variability. 

We define the cost stabilizers pay for niche construction as the extent of change to the environmental switching rate, or the KL divergence between the natural environmental rate $1/\te$ and the time-averaged, modified rate $\br{1/\tildete}$,
\begin{align}
\begin{aligned}
	G(1/\te, \br{1/\tildete}) &= \frac{1}{\te} \log_2 \left( \frac{1/\te}{\br{1/\tildete}} \right) + \\
		&\qquad\left(1-\frac{1}{\te}\right) \log_2 \left( \frac{1-1/\te}{1-\br{1/\tildete}} \right).\label{eq:G}
\end{aligned}
\end{align}
The quantity $G$ depends implicitly on stabilization strength $\alpha$ because smaller $\alpha$ slows the environment further. For passive agents and destabilizers, $G=0$ by definition because non-stabilizers fit to $\te$ and only stabilizers benefit from the slower timescale with monopolization.

We finally consider the cost of precision, which we assume to be given by the information obtained by the agent from sampling the environment,
\begin{align}
	C(\tc) &= \log_2 \tc.\label{eq:sensory cost}
\end{align} 
Sensory complexity means that higher precision implies higher expenditure to obtain such precision, given by the KL divergence between environment configuration and agent behavior, $C \sim -\log_2(\sigma^2)$ leaving out constants. This depends on the variance of agent measurement noise $\sigma^2 = \pe(s,t)[1-\pe(s,t)] / \tc$. Infinitely precise sensory cells lead to diverging cost, whereas imprecise cells are cheap.

Putting these costs together with divergence $\bar{D}$, we obtain the total divergence
\begin{align}
	\mathcal{D} &= \bar{D} + \mu H + \chi G + \beta C.\label{eq:total divergence}
\end{align}
Weights $\mu\geq0$, $\chi\geq0$, $\beta\geq0$ represent the relative contribution of these costs. As a result, we can distinguish dominant strategies by comparing total divergence such as between the pair of destabilizer and stabilizer strategies shown in Figure~\ref{gr:fitness}. Large $\mu$, or high complexity cost, means that a pure population of stabilizers would be stable to invasion from destabilizers. Whereas for large $\chi$, or heavy stabilization cost, the opposite is true.  The generalized measure of adaptive cost in Eq~\ref{eq:total divergence}, given the weights, carves out regions of agent morphospace along axes of computational cost. This is a morphospace that captures the relative advantage of internal versus external memory that can be thought of as a space of evolutionary outsourcing. 


As has often been remarked in relation to evolution, survival is not the same as arrival.  We now determine when stabilizer strategies can emerge in this landscape. 
We start with a pure population of passive agents with stabilization strength $\alpha=0$ and poised about optimal memory duration $\ta=\ta^*$ determined by minimizing both divergence $\bar{D}$ and complexity $\mu H$. Whether or not stabilizers emerge under mutation and selection can be determined through adaptive dynamics \cite{brannstromHitchhikerGuide2013,dieckmannCoevolutionaryDynamics,dieckmannDynamicalTheory1996}, that is by inspecting the gradient of the total divergence along the parameters $(\partial_{\ta} \mathcal{D}, \partial_{\alpha} \mathcal{D}, \partial_{\tc} \mathcal{D})$, or memory complexity, stabilizer strength, and precision. As we show in SI Appendix~\ref{si sec:evolution} and Eq~\ref{si eq:gradient}, the gradient terms can be calculated under a set of perturbative approximations. 
Using local convexity about optimal memory $\ta^*$, we show that the term $\partial_\alpha \mathcal{D}$ drives passive agents to smaller $\alpha$ and slower timescales; it originates from combining the scaling law from Eq~\ref{eq:D scaling} and complexity of memory. 
The term $\partial_{\tc}\mathcal{D}$ shows that precision tends to decrease when the cost gradient $\partial_{\tc}(\beta C)$ dominates over $\partial_{\tc}\bar D$. In this case, the general conditions $\partial_\alpha \mathcal{D}<0$ and $\partial_{\tc}\mathcal{D}<0$ funnel a passive population towards stabilization and reduced precision. 

\section*{Discussion}
Life is adaptive, but optimal adaptation would seem to depend on a multitude of properties of both organism and environment, which have been studied in a wide literature (Table~\ref{tab:refs}). To the contrary, we predict that it does not. This becomes clear once we organize crucial aspects of adaptation into a unified framework in terms of timescales including niche construction that speeds up or slows down the environment (Figure~\ref{gr:overview}). We find that memory duration, under a wide range of assumptions and conditions, scales sublinearly with environmental rates of change (Figure~\ref{gr:tau scaling}). This essentially derives from the competition between using current but noisy information and the reliance on outdated but precise information, leading to a universal, optimal timescale for adaptation. 
Importantly, sublinear scaling implies that persistent features of the environment can be more efficiently encoded the longer-lasting they become; there is an economy of scale.

Yet, memory remains costly as it requires investing in neural tissue. To estimate this cost and how it might affect adaptation, we use metabolic scaling theory to estimate how much neural tissue an organism must allocate to memory for a given rate of environmental change. We find that the metabolic costs of memory can increase super-linearly with the persistence time of environmental statistics. Thus, while memory need not grow in proportion to environmental stability, costs of memory could increase disproportionately (Figure~\ref{gr:metabolism}). Because adaptive costs peak at short timescales, this suggests that adaptive costs are most important for organisms with short lifespans such as insects.

When the costs of adaptation are greater than the metabolic costs of memory, active modification of the environment such as stabilizing niche construction can be favored. In this case the organism intervenes on the environmental timescale to decrease volatility. Although outsourcing of memory to the environment reduces the organism's need to adapt, it introduces two new problems. First, active modification is itself not free. Second, slow environmental variables created by active modification are public goods that can be exploited by free riders.

To address the costs of active modification and free riding, we introduce game dynamics considering the information costs of adaptation including the complexity of memory. Unlike memory duration, memory complexity quantifies the effective number of states that agents occupy. Starting with passive agents, we find that the spontaneous emergence of adaptive dynamics stabilizes the environment, lengthening the optimal memory duration $\ta^*$ and thereby making weak stabilizers less competitive. This moves a population as a whole towards slower timescales. In other words, stabilizing niche construction, because of the economy of scale with respect to memory, requires proportionally less neural tissue for memory relative to the size of the whole brain as given by metabolic scaling theory. This is effectively outsourcing memory from neural tissue to the environment. As a possible consequence, organisms could reduce absolute brain size or invest in a larger behavioral repertoire, increasing competitiveness by monopolizing a larger number of environmental states. Do learning agents in volatile environments ``given a choice'' to invest in additional memory or to directly change the environment favor the latter?

This hypothesis is consistent with related work on institutions and social structure as a form of collectively encoded memory \cite{poonInstitutionalDynamics2022, brushConflictsInterest2018, ramos-fernandezCollectiveComputation2020, leeCollectiveMemory2017} or as devised constraints (e.g.~reference \citenum{northUnderstandingProcess2005}) that slow down the need to acquire functional information. In pigtailed macaque society (reviewed in reference \citenum{flackCoarsegrainingDownward2017}), individuals collectively compute a social-power distribution from status signaling interactions. The distribution of power as a coarse-grained representation of underlying fight dynamics changes relatively slowly and consequently provides a predictable social background against which individuals can adapt. By reducing uncertainty and costs, the power distribution facilitates the emergence of novel forms of impartial conflict management. Conflict management, in turn, further reduces volatility, allowing individuals to build more diverse and cohesive local social niches and engage in a greater variety of socially positive interactions \cite{Flack:2006fk}. In other words, outsourcing memory, in this case, of fight outcomes, to a stable social structure in the power distribution allows for a significant increase in social complexity. More generally, we anticipate that one of the features of slowing environmental timescales, including social environments fostered by institutions, might the emergence of new functions \cite{flackLifeInformation2017a}.

Without feedforward and feedback loops between environment and agent such as in the case of the passive agent, our framework is akin to the classical problem of learning. This has been a major problem of interest in foraging \cite{mcnamaraOptimalForaging1985}, neural circuits that adapt to changing input distributions \cite{brennerAdaptiveRescaling2000,gershmanNeuralCosts2010,foxBayesianNonparametric2011} and modes of prediction in order to best adapt to multiple clustered sets of statistics \cite{gershmanNeuralCosts2010,gershmanStatisticalComputations2014}. We introduce here a minimal modeling framework for connecting learners to active agents that modify the environment through the act of adaptation. Our framework provides a first-order approximation to this extended space, which could itself be extended in several directions to include how agents physically modify their environments, connecting to the physics of behavior with the physics of information \cite{leeDiscoveringSparse2022}.

\matmethods{
The code used to generate these results will be made available on GitHub at \url{https://github.com/eltrompetero/adaptation}.

\subsection*{Numerical solution to model}
Given Eqs~\ref{eq:field}--\ref{eq:memory} defining the binary agent, we calculate agent behavior in two ways. The first method is with agent-based simulation (ABS). We generate a long time series either letting the environment fluctuate independently and training the agent at each moment in time or coupling environmental fluctuations at each time step with the state of the agent. By sampling over many such iterations, we compute the distribution over agent bias given environmental bias, $q(h|\he)$, which converges to a stationary form.

This principle of stationarity motivates our second solution of the model using an eigenfunction method. If the distribution is stationary, then we expect that under time evolution that the conditional agent distribution map onto itself
\begin{align}
	q(h|\he) &= \mathcal{T}[q(h|\he)]. 
\end{align}
If the time-evolution operator $\mathcal T$ evolves the distribution over a single time step, the external field can either stay the same with probability $1-1/\te$ or reverse with probability $1/\te$. 

For either for these two possible alternatives over a single time step, we must convolve the distribution with the distribution of noise for the field $\eta_\tc$. The distribution of noise derives from agent perceptual errors $\epsilon_\tc$ on the estimated probabilistic bias of the environment (Eq~\ref{eq:bias}). Hence, the corresponding error distribution for the bias $\eta_\tc$ originates from the binomial distribution through a transformation of variables. We can simplify this because in the limit of large sensory cell sample size $\tc$ the binomial distribution converges to a Gaussian and a concise representation of the distribution of $\eta_\tc$ becomes accurate. Using Eq~\ref{eq:field}, we find that the distribution of perceptual errors in the bias yields
\begin{align}
\begin{aligned}
	\rho(\eta_\tc,t) &= (8\pi \sigma^2)^{-1/2} \exp\big\{ -\left[ \tanh \he(t) -\right. \\
	& \left.\left.\tanh (\he(t) + \eta_\tc) \right]^2/8\sigma^2 \right\}\sech^2\left(\he(t)+\eta_{\tc}\right).
\end{aligned}\label{eq:field error distribution}
\end{align}
Here, the agent's perceptual estimate of the environment includes finite-sample noise determined by the sensory cell precision $1/\tc$. At finite $\tc$, there is the possibility that the agent measure a sample from the environment of all identical states. In our formulation, the fields then diverge as do the fields averaged over many separate measurements. We do not permit such a ``zero-temperature'' agent that freezes in a single configuration in our simulation just as thermodynamic noise imposes a fundamental limit on invariability in nature. Our agents inhabit an {\it in silico} world, where the corresponding limit is fixed by the numerical precision of the computer substrate, so we limit the average of the bits sampled from the environment to be within the interval $[-1+10^{-15},1-10^{-15}]$. This is one amongst variations of this idea that inference is constrained by regularization, Bayesian priors, Laplace counting (in the frequentist setting), etc. Regardless of the particular approach with which finite bounds might be established, they are only important in the small $\tc$ limit. See SI Appendix~\ref{sec:abm solution}.

Given the Gaussian approximation to precision error, we propagate the conditional distribution over a single time step, defining a self-consistent equation that can be solved by iterated application. To make this calculation more efficient, we only solve for abscissa of the Chebyshev basis in the domain $\beta\in[0,1]$, fixing both the endpoints of the interval including the exact value for $\beta=1$ from Eq~\ref{eq:divergence at infty} \cite{presswilliamh.NumericalRecipes2007} (more details in SI Appendices~\ref{sec:abm solution} and \ref{si sec:eigensolution}). In Figure \ref{gr:cond prob check}, we show that our two methods align for a wide range of agent memory $\ta$. Importantly, the eigenfunction approach is much faster than ABS for large $\tc$ because the latter can require a large number of time steps to converge. On the other hand, ABS is relatively fast for small $\tc$. Thus, these two approaches present complementary methods for checking our calculation of agent adaptation.

\subsection*{Divergence curves}
To measure how well agent behavior is aligned with the environment, we compare environment $\pe(s,t)$ and agent $p(s,t)$ with the KL divergence at each time step to obtain the agent's typical loss in Eq~\ref{eq:D}. Equivalently, we can average over the stationary distribution of fields conditional on environment
\begin{align}
	\bar D &= \frac{1}{N_E}\sum_E \int_{-\infty}^\infty dh\, q(h|\he) D_{\rm KL}[\pe(\he)||p(h)],\\
\intertext{where we sum over all possible environments $E$ and weight them inversely with the number of total environments $N_E$. For the binary case, $N_E=2$. We furthermore simplify this for the binary case as}
	\bar D &= \int_{-\infty}^\infty dh\, q(h|\he=h_0) D_{\rm KL}[\pe(\he=h_0)||p(h)]. \label{eq:D stationary form}
\end{align}
In Eq~\ref{eq:D stationary form}, we have combined the two equal terms that arise from both positive $\he=h_0$ and negative $\he=-h_0$ biases of the environment.

In Figure~\ref{gr:tau scaling}A and B, we show divergence as a function of agent memory over a variety of environments of varying correlation time $\bar D(\ta,\te)$. When the agent has no memory, its behavior is given solely by the properties of the sensory cells as is determined by the integration time $\tc$. Then, we only need account for the probability that the environment is in either of the two symmetric configurations and how well the memoryless agent does in both situations. Since the configurations are symmetric, the divergence at zero memory is
\begin{align}
\begin{aligned}
	\bar D(\ta=0) &= \int_{-\infty}^{\infty} d\eta_\tc\,\rho(\eta_\tc|\he=h_0) \times\\
		&\sum_{s\in\{-1,1\}} \pe(s|\he=h_0) \log_2\left( \frac{\pe(s|\he=h_0)}{p(s)}\right),
\end{aligned}\label{eq:D at 0}
\end{align}
where the biased distribution of environmental state $\pe$ and the error distribution $\rho$ from Eq~\ref{eq:field error distribution} are calculated with environmental bias set to $\he=h_0$. Note that this is simply Eq~\ref{eq:D stationary form} explicitly written out for this case. 

At the limit of infinite agent memory, as in the right hand side of Figure~\ref{gr:tau scaling}A, passive agents have perfect memory and behavior does not budge from its initial state. Assuming that we start with an unbiased agent such that $q(h) = \delta(h)$, the Dirac delta function, the agent's field is forever fixed at $h=0$. Then, divergence reduces to
\begin{align}
\begin{aligned}
	\bar D(\ta=\infty) &= 1 - S[\pe],
\end{aligned}\label{eq:divergence at infty}
\end{align}
where the conditional entropy $S[\pe] = -\pe(s|h=h_0)\log_2\pe(s|h=h_0) -[1-\pe(s|h=h_0)]\log_2[1-\pe(s|h=h_0)]$.

\subsection*{Scaling argument for optimal memory}\label{sec:scaling details}
As is summarized by Eq~\ref{eq:tau scaling trade-off}, the value of optimal memory can be thought of as a trade-off between the costs of mismatch with environment during the transient adaptation phase and gain from remembering the past during stable episodes. In order to apply this argument to the scaling of divergence, we consider the limit where the environment decay time $\te$ is very long and agent memory $\ta$ is long though not as long as the environment's. In other words, we are interested in the double limit $\ta\rightarrow\infty$ and $\ta/\te\rightarrow0$. Then, it is appropriate to expand divergence in terms of the error in estimating the bias
\begin{align}
\begin{aligned}
	\bar D &= \Bigg\langle\sum_{s\in\{-1,1\}} \pe(s,t)\log\pe(s,t) - \\
	&\qquad\qquad\pe(s,t)\log[\pe(s,t)+\epsilon_\tc(t)]\Bigg\rangle,
\end{aligned}\label{eq:D exp1}
\end{align}
where the average is taken over time. Considering only the second term and simplifying notation by replacing $\epsilon_\tc(t)$ with $\epsilon$,
\begin{align}
\begin{aligned}
	&\br{\pe(s,t)\log\pe(s,t) + \log[1+\epsilon/\pe(s,t)]}\\
	&\qquad\approx\br{\pe(s,t)\log\pe(s,t) +\frac{\epsilon}{\pe(s,t)} - \frac{1}{2}\frac{\epsilon^2}{\pe(s,t)^2}},
\end{aligned}\label{eq:D exp2}
\end{align}
where the average error $\br{\epsilon}=0$ and assuming that the next nontrivial correlation of fourth order $\mathcal{O}\left(\br{\epsilon^4}\right)$ is negligible. Plugging this back into Eq~\ref{eq:D exp1},
\begin{align}
	\bar D &\approx \sum_{s\in\{-1,1\}} \frac{\te-\ta}{\te}\frac{\br{\epsilon^2}}{\pe(s)^2} + \frac{\ta}{\te}\br{\frac{\epsilon^2}{\pe(s,t)^2}}. \label{eq:D exp3}
\end{align}
The first term in Eq~\ref{eq:D exp3} relies on the fact that when environmental timescales are much longer than agent memory, the errors become independent of the state of the environment. Thus, we can average over the errors separately, and the environment configuration average can be treated independently of time $\pe(s,t)\rightarrow\pe(s)$. The second term, however, encases the transient dynamics that follow immediately after a switch in environmental bias while the agent remembers the previous bias. It is in the limit $\ta/\te\rightarrow0$ that we can completely ignore this term and the scaling for optimal memory $\ta^*\sim\te^{1/2}$ from Eq~\ref{eq:tau*} is the relevant limit that we consider here.

Since the errors with which the agent's matching of environmental bias is given by a Gaussian distribution of errors, the precision increases with the number of samples taken of the environment: it should increase with both sensory cell measurement time $\tc$ as well as the typical number of time steps in the past considered, $\ta=-1/\log\beta$.  Thus, we expect the scaling of divergence at optimal memory to be
\begin{align}
	\bar D^* \sim \frac{1}{\ta^*\tc},\label{eq:Dstar scaling}
\end{align}
which with Eq~\ref{eq:tau*} leads to the scaling of optimal memory with environment decay time Eq~\ref{eq:D scaling}. Though the scaling with precision timescale $\tc$ in Eq~\ref{eq:Dstar scaling} is at $\ta=\ta^*$, it is clear that a similar scaling with $\tc$ holds at $\ta=0$, where only precision determines divergence. However, such a scaling does not generally hold for any fixed $\ta$, the trivial case being at $\ta=\infty$, where divergence must go to a constant determined by environmental bias.
}

\showmatmethods{} 

\acknow{E.D.L.~was supported by the Omega Miller Program at the Santa Fe Institute. D.C.K.~and J.F.~are grateful for support from the James S.~McDonnell Foundation 21st Century Science Initiative-Understanding Dynamic and Multi-scale Systems.}

\showacknow{} 

\clearpage
\appendix
\renewcommand\thefigure{S\arabic{figure}}  
\setcounter{equation}{0}
\renewcommand{\theequation}{S\arabic{equation}}
\renewcommand\thetable{S\arabic{table}}

\section{Agent-based simulation}\label{sec:abm solution}
To complement the eigenfunction solution described in Appendix~\ref{si sec:eigensolution}, we present the agent-based simulation.

After having specified the environmental bias $\he(t)$, we generate a sample of $\tc$ binary digits from the distribution $\pe(s,t)$. From this sample, we calculate the mean of the environment $\br{s}$ which is bounded in the interval $[-1+10^{-15}, 1-10^{-15}]$. These bounds are necessary to prevent the measured field $\hat h(t)$ from diverging and reflects the fact that {\it in silico} agents have a finite bound in the values they can represent, mirroring finite cognitive resources for biological or social agents as discussed in Materials and Methods. We combine this estimated field $\hat h(t)$ with the one from the aggregator having set the initial value condition $H(0)=0$. Given the estimate of the field $h(t)$, we compute the Kullback-Leibler (KL) divergence between the agent distribution $p(s)$ and the environment $\pe(s)$. 

When we calculate the divergence landscape across a range of different agent memories, we randomly generate the environment using the same seed for the random number generator. Though this introduces bias in the pseudorandom variation between divergence for agents of different types, it makes clearer the form of the divergence landscape by eliminating different offsets between the points. Our comparison of this approach with the eigenfunction solution in Appendix~\ref{si sec:eigensolution} provides evidence that such bias is small with sufficiently long simulations. For the examples shown in the main text, we find that total time $T=10^7$ or $T=10^8$ are sufficient for convergence to the stationary distribution after ignoring the first $t=10^4$ time steps.

\section{Eigenfunction solution}\label{si sec:eigensolution}
We present more details on top of those in Materials and Methods on the iterative, eigenfunction solution to the divergence of an agent relying on the fact that the distribution of agent bias $q(h)$ becomes stationary at long times.

Let us first consider the case of the passive agent. After sufficiently long time, the distribution of agent behavior $q(h)$ and the distribution conditioned on the two states of the environment $q(h|\he=h_0)$ and $q(h|\he=-h_0)$ converge to stationary forms. Assuming that the distributions have converged, we evolve the distribution a single time step. If the external field $\he(t)=h_0$, then it either stays fixed with probability $1-1/\te$ or it switches to the mirrored configuration with probability $1/\te$. 

Considering now the evolution of the conditional probability $q(h|\he=h_0)$, we note that the state of the agent will be either be convolved by the distribution of sampling error at the next time step or lose probability density from a switching field. Since we are considering a symmetric configuration, however, the mirrored conditional density will reflect the same probability density back such as in Eq~\ref{eq:rho 1}. Thus, Eq~\ref{eq:rho 1} is satisfied by the conditional density of agent bias that is solved by the eigenfunction for $q(h|\he)$ with eigenvalue $1$. By the Perron-Frobenius theorem when considering normalized eigenvectors, this is the unique and largest eigenvalue that returns the stationary solution.

To extend this formulation to active agents, we must also account for the dependence of the rate of switching on the distance between agent and environmental bias. This additional complication only requires a modification of Eq~\ref{eq:rho 1} to include such dependence in the rate coefficients. Thus, all types of agents can be captured by this eigenfunction solution and solved by iteration til convergence.

\begin{figure*}[tb]\centering
	\includegraphics[width=.35\linewidth]{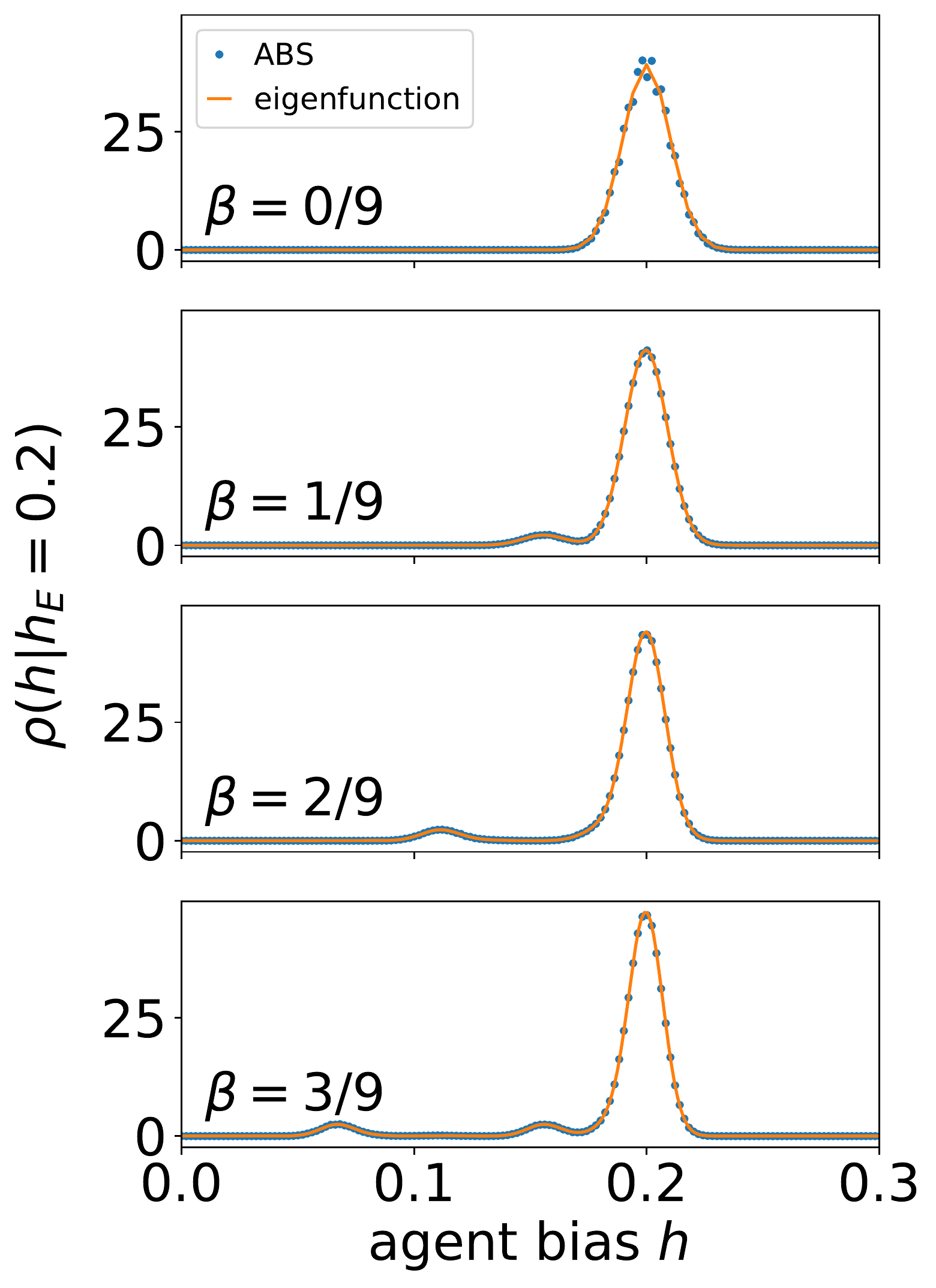}\includegraphics[width=.35\linewidth]{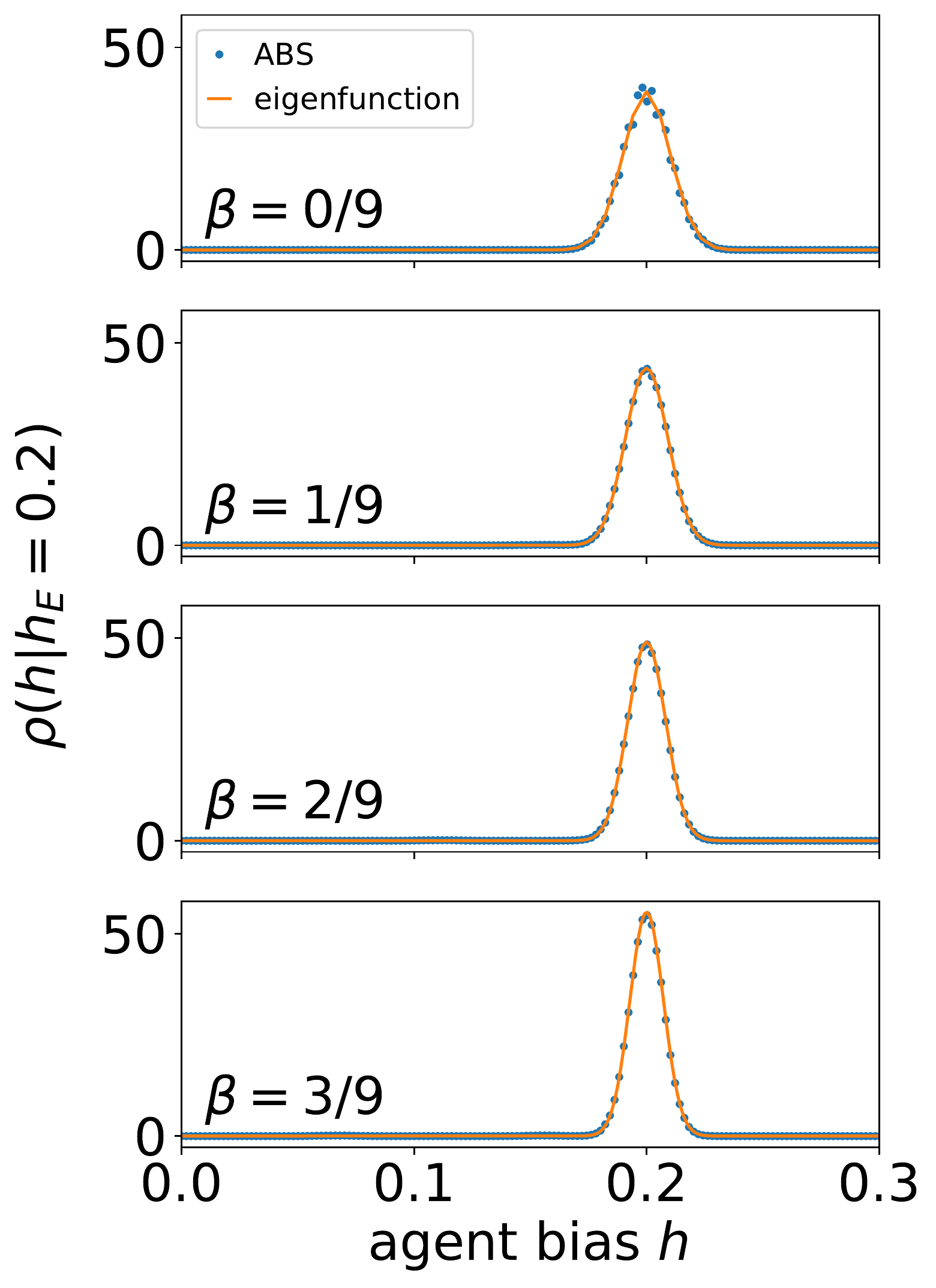}
	\caption{Comparison of agent-based simulation (ABS) and eigenfunction solution for the conditional probability distribution of agent bias $q(h|h_0)$ for (left) a passive agent and (right) stabilizer. Agent-based simulation returns a normalized histogram that aligns closely with the eigenfunction solution. Environment timescale $\te=20$ and bias $h_0=0.2$. Spacing of discrete domain in eigenfunction solution determined in proportion with typical width of the peak around $h=h_0$, which scales as in Eq~\ref{eq:field error distribution} and inversely with the square root of agent memory $\ta$.}\label{gr:cond prob check}
\end{figure*}

Eq~\ref{eq:rho 1} is only independent of time when agent memory $\ta=0$. When there is finite memory, or $\beta>0$, the distribution $q(h,t)$ ``remembers'' the previous state of the environment such that we must iterate Eq~\ref{eq:rho 1} again. Over many iterations, we will converge to the solution, but the convergence slows with agent memory which introduces ever slower decaying eigenfunctions. An additional difficult arises because the narrowing in the peak of the agent's estimate of the environment, like the peaks shown in Figure~\ref{gr:cond prob check}, require increased numerical precision. As a result, increasing memory and computational costs make it infeasible to calculate the eigenfunction with high precision for $\beta$ close to 1.

\begin{figure}[tb]\centering
	\includegraphics[width=.9\linewidth]{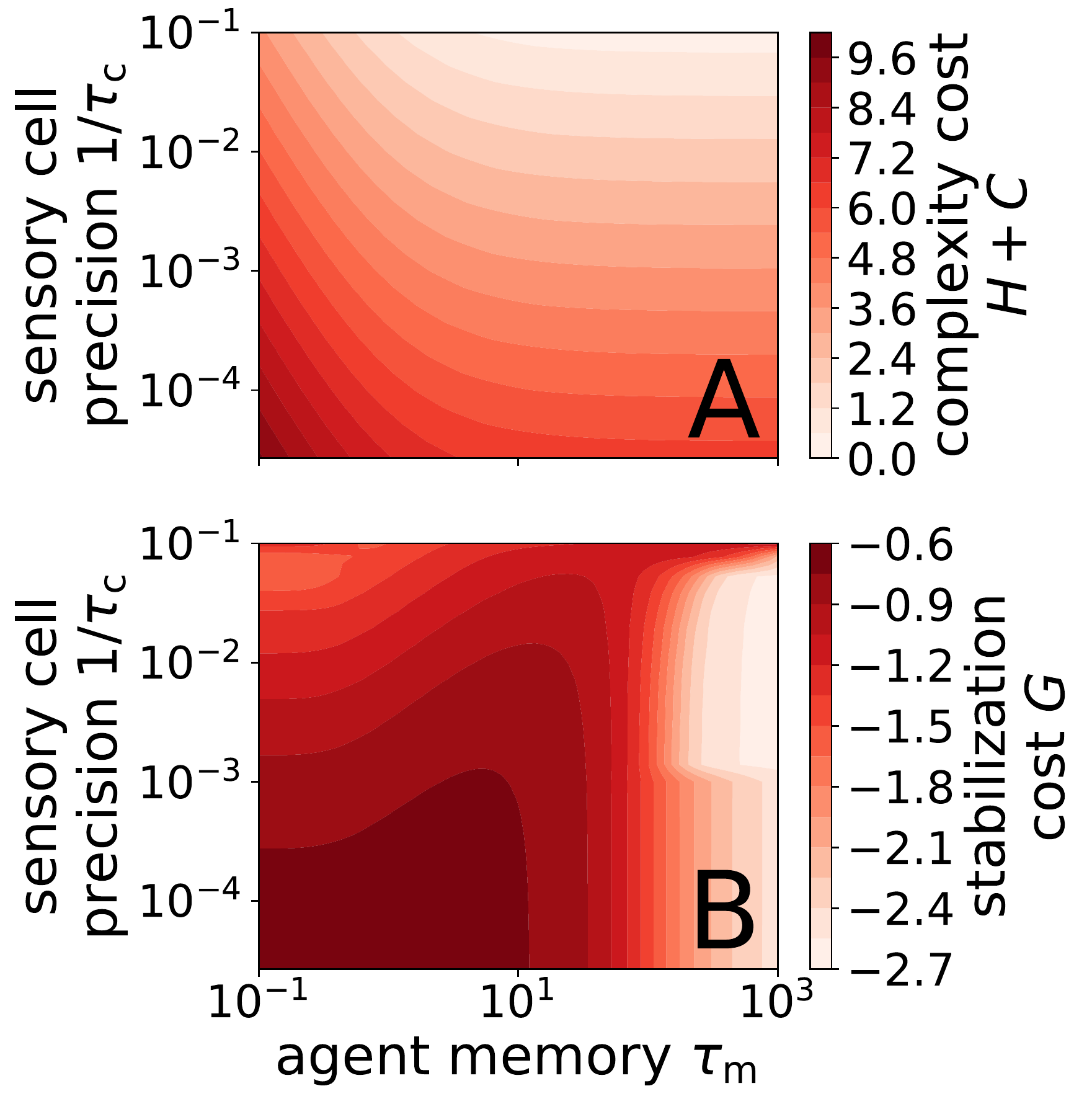}
	\caption{Landscape of the costs we consider as (A) a combined agent complexity and (B) a stabilization cost. (A) Isocontours defined as sum of memory complexity and sensory precision costs. The values have been offset to ensure that the costs are positive over the shown landscape calculated from memory (Eq~\ref{eq:memory cost}) and sensory cost (Eq~\ref{eq:sensory cost}). (B) As memory $\ta\rightarrow\infty$, stabilization cost converges to a finite value that can be calculated exactly from noting that agent behavior has probability density fixed at its starting point, $q(h)=\delta(h)$. A kink in the contours at $1/\tc=10^{-3}$ arises from numerical precision errors where we matched up ABS and eigenfunction methods.}\label{gr:costs}
\end{figure}

\begin{figure}[tb]\centering
	\includegraphics[width=.85\linewidth]{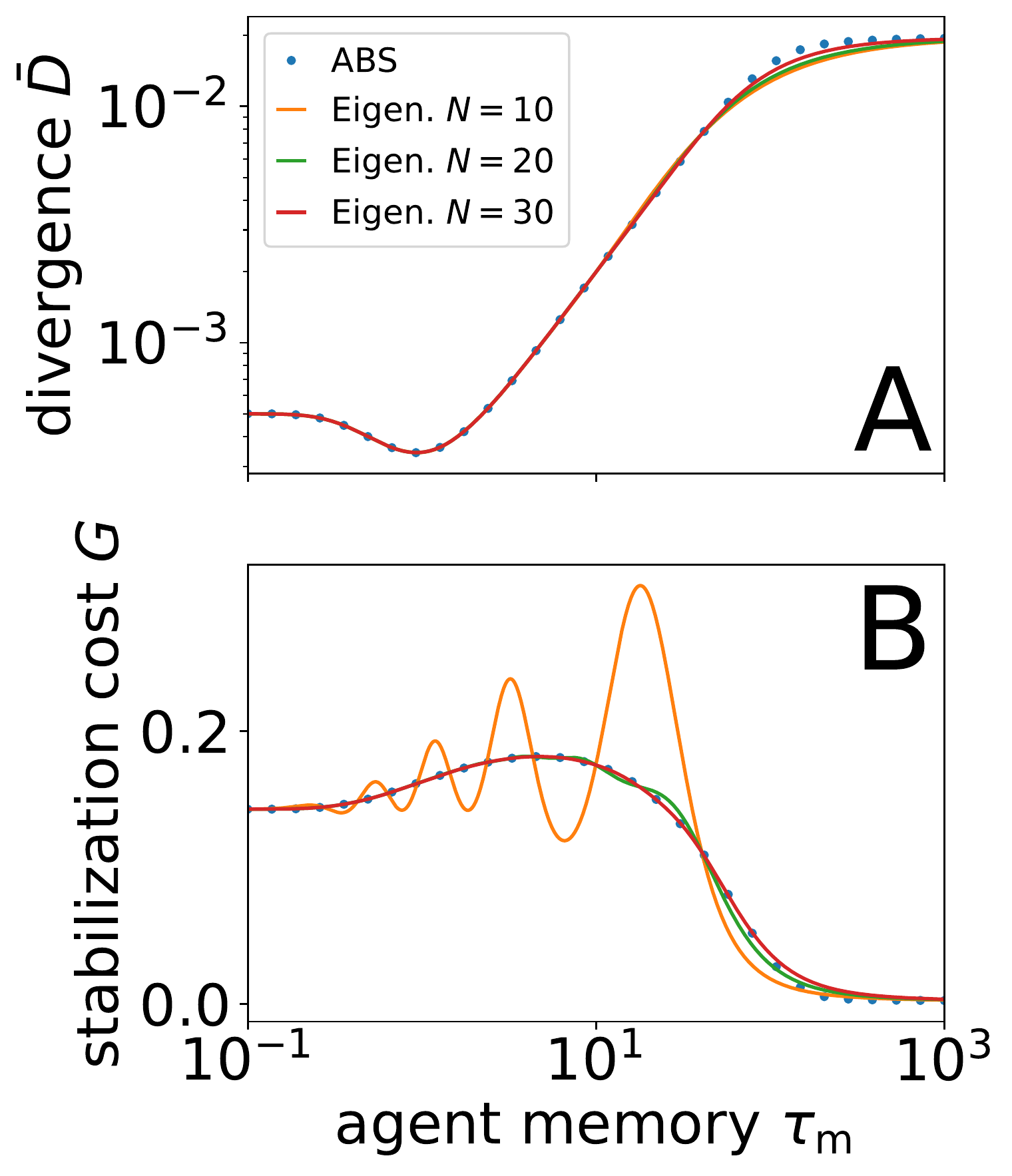}
	\caption{Example of convergence of least-squares fit of Chebyshev polynomial with increasing number of abscissa $N$ with the eigenfunction solution. (top) For comparison, divergence $D$ as calculated from the agent-based simulation (ABS). The eigenfunction solution is close even with a relatively small number of points fit to a 9th-degree Chebyshev polynomial. Both methods are especially effective when the environmental timescale is small as is here, where $\te=10$. The bias $h_0=0.2$. (bottom) Stabilization cost is similarly interpolated, but it is slower to converge with visible oscillations disappearing by $N=30$. For $N=20$ and $N=30$, not all the points fell within the convergence criterion and only 19 and 28 points were fit, respectively. For both plots, the Chebyshev polynomial approximation is slowest to converge near the sharp bends at large $\ta$. ABS is run for $10^7$ time steps.}\label{gr:chebyshev convergence}
\end{figure}

Instead of calculating the full functional form directly below but not at the limit $\beta\rightarrow1$, we use the output of the iterative eigenfunction procedure as input for an interpolation procedure using Chebyshev polynomials. We iterate Eq~\ref{eq:rho 1} for $\beta$ equal to the Gauss-Lobatto abscissa of the Chebyshev polynomial of degree $d$, mapping the interval $\beta\in[0,1]$ to the domain $x\in[-1,1]$ for the set of Chebyshev polynomials \cite{presswilliamh.NumericalRecipes2007}. The Gauss-Lobatto points include the endpoints $\beta=0$ and $\beta=1$, the first of which is trivial numerically and the latter for which we have an exact solution given in Eq~\ref{eq:divergence at infty}. Then, we exclude calculated values for large $\beta$ that show large iteration error $\epsilon>10^{-4}$. This threshold, however, leaves the coefficients of the Chebyshev polynomial undetermined. We instead interpolate these remaining $N-k$ points by by fitting a Chebyshev polynomial of degree $N-k-1$ with least-squares on the logarithm of the divergence. A similar procedure can be run for the stabilization cost from Eq~\ref{eq:G} to obtain Figure~\ref{gr:costs}B. We find that typically $N=30$ or $N=40$ starting abscissa with a maximum of $10^3$ iterations are sufficient to obtain close agreement with the agent-based simulation (ABS) from Appendix~\ref{sec:abm solution} (Figure~\ref{gr:chebyshev convergence}). This interpolation procedure does not work well with ABS because small stochastic errors can lead to high-frequency modes in interpolation (and thus large oscillations), errors that can be essentially driven to zero exponentially fast for the eigenfunction method.


\begin{widetext}
\begin{align}
\begin{aligned}
	q(h,t|\he=h_0) &= \left(1-\frac{1}{\te}\right) \int_{-\infty}^\infty\int_{-\infty}^\infty \rho(\eta_\tc|\he=h_0)q(h,t-1|\he=h_0)\delta(h-h_0-\eta_\tc)\,d\eta_\tc\,dh +\\
	&\frac{1}{\te} \int_{-\infty}^\infty\int_{-\infty}^\infty \rho(\eta_\tc|\he=-h_0)q(h,t-1|\he=-h_0)\delta(h+h_0-\eta_\tc)\,d\eta_\tc\,dh.
\end{aligned}\label{eq:rho 1}
\end{align}
\end{widetext}

\begin{figure}[tb]\centering
	\includegraphics[width=\linewidth]{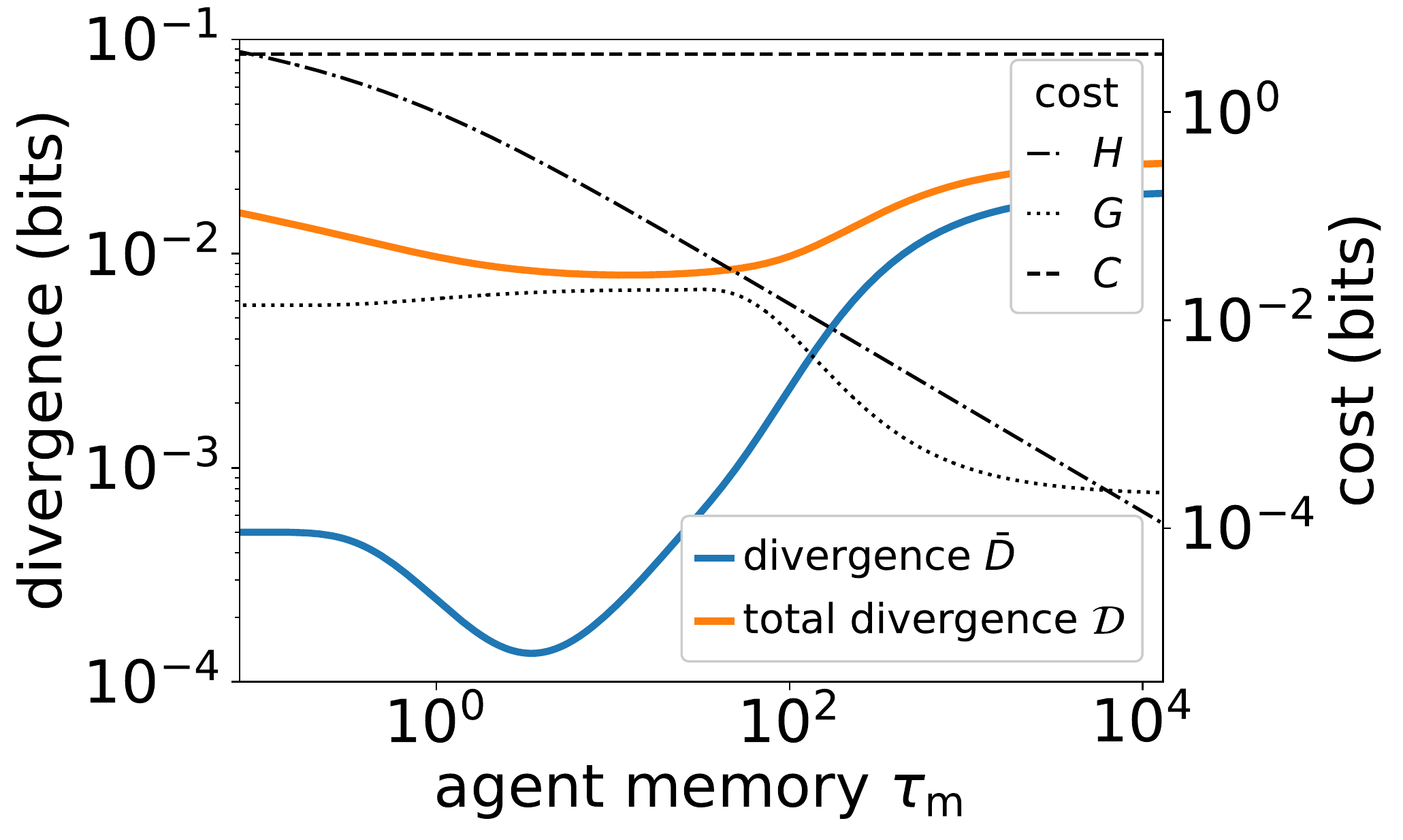}
	\caption{Example of cost functions for stabilizers with varying memory but fixed sensory precision. (blue) Without costs, divergence profile shows only a single global minimum. (orange) With costs, we obtain degenerate minima at memory values around $\ta=0$ and $\ta=20$. Eigenfunction solution parameters specified in Materials and Methods code.}\label{gr:cost trade-off ex} 
\end{figure}

\section{Evolution of reduced complexity}\label{si sec:evolution}
We consider a population of passive agents, or an agent with stabilization parameter $\alpha=0$, precision timescale $\tc$, and optimal memory $\ta^*$, the variables that determine agent fitness. Assuming that the canonical equation for evolution applies (i.e.~mutations only change phenotype and fitness slightly, the population dynamics move much faster than the evolutionary landscape such that we can assume a single phenotype dominates), the rate at which the population evolves across the phenotypic landscape is proportional to the fitness gradient. In addition to this assumption, we will assume that the population is always poised at optimal memory, an assumption that will be made clear below.

We recall that the total divergence consists of the time-averaged divergence $\bar{D}$, statistical complexity cost $H$, stabilization cost $G$, and precision cost $C$
\begin{align}
	\mathcal{D} &= \bar{D} + \mu H(\ta) + \chi G(\te,\tildete) + \beta C(\tc)
\end{align}
with semi-positive weights $\mu$, $\chi$, and $\beta$. In Figure~\ref{gr:cost trade-off ex}, we show each the divergence of a stabilizer without such costs in blue, each of these costs separately in black, and their sum in orange to generate the total divergence in Eq~\ref{eq:total divergence}. For the evolutionary dynamics, we must calculate the gradient $(\partial_{\ta}\mathcal{D}, \partial_{\alpha}\mathcal{D}, \partial_{\tc}\mathcal{D})$ determining the evolution in the properties of the agent. We calculate these term by term and then put them together at the end.

We assume that agent memory $\ta$ is at the minimum of the combination of time-averaged divergence $\bar D$ and statistical complexity cost $\mu H$ (stabilization is zero for passive agents). Since divergence has a unique minimum and complexity monotonically approaches $H(\ta=\infty)=0$, the addition of complexity only shifts optimal memory to a larger value. Without the complexity cost, we have that small deviations about optimal memory can be represented by a quadratic function for some positive constant $a$,
\begin{align}
	\bar{D} &= D^* + a (\ta-\ta^*)^2,\\
	\intertext{where we write}
	D^* &= D_0(\ta^*)^{1/2}
\end{align}
for some positive constant $D_0$. Once we have accounted for a perturbative addition from memory complexity, however, we have a shifted optimal memory
\begin{align}
\begin{aligned}
	\ta^{**} &= \ta^* +\frac{\mu}{2(\log2)a\ta^*(\ta^*+1)} + \mathcal{O}(\mu^2) \label{si eq:tss}
\end{aligned}
\end{align}
obtained from $\partial_{\ta}[\bar{D} + \mu H]=0$ and using the approximation that $\mu$ is small. Eq~\ref{si eq:tss} shows us that memory complexity, the term proportional to $\mu$, drives optimal memory $\ta^{**}$ up.

Taking the approximation in Eq~\ref{si eq:tss} the shifted optimal divergence, denoted by an apostrophe, becomes
\begin{align}
\begin{aligned}
	\bar{D}'(\ta^{**}) &= D^* + a\frac{\mu^2}{4(\log2)^2(\ta^*)^2(\ta^*+1)^2} + \mathcal{O}(\mu^3).
\end{aligned}
\end{align}
Again, perturbations about the local optimum lead to
\begin{align}
\begin{aligned}
	\bar{D}'(\ta) &\approx D^* + a\frac{\mu^2}{4(\log2)^2(\ta^*)^2(\ta^*+1)^2} + \\
	& \qquad b(\ta-\ta^{**})^2\label{si eq:shifted div}
\end{aligned}
\end{align} 
for some positive constant $b$, which implicitly depends on the complexity cost. Eq~\ref{si eq:shifted div} expresses local convexity about shifted optimal memory $\ta^{**}$ according to the corresponding shifted divergence $\bar{D}'$. This indicates how the population is poised along the ridge of optimal memory given a perturbative cost of memory complexity. 

Then, time-averaged divergence will grow because optimal memory changes. Assuming that the population is at optimal memory, we obtain for the partial derivative with respect to $\alpha$
\begin{align}
\begin{aligned}
	\partial_\alpha \bar{D}' &= \left[\frac{D_0}{2} (\ta^*)^{-\frac{3}{2}} + a\frac{\mu^2(2\ta^*+1)}{2(\log2)(\ta^*)^3(\ta^*+1)^3} \right] \left|\frac{\partial \ta^*}{\partial \alpha}\right|,\\
\end{aligned}\label{si eq:D star}
\end{align}
where we have used the fact that optimal memory must increase with stronger stabilizer, or that $\partial_\alpha\ta^*<0$, to explicitly pull out a negative sign. Given that we are in the scalin regime, this confirms that in Eq~\ref{si eq:D star} divergence at optimal memory decreases as $\alpha$ approaches $-1$ from above as expected.

Niche-constructing stabilization changes the environmental timescale through feedback. We start by considering over a long period of time the average over many environmental switches
\begin{align}
\begin{aligned}
	\br{1/\tildete} &= 1/\te + \alpha \br{\frac{v^2}{v^2 + (h-\he)^2}},\\
		&= 1/\te + \alpha f(\ta).
\end{aligned}\label{si eq:tildete1}
\end{align}
Since we do not know the exact form of the second term on the right hand side, we represent it as some function $f$ that represents an average over time. For notational simplicity, we only make explicit $f$'s dependence on $\ta$, but it depends on agent properties and environmental timescale. Now, a change in $\alpha$ also indirectly affects $\ta^*$ because the environmental timescale will change, reducing or increasing the agents ability to track the new environment. For example, with the passive agent, an increase in $\alpha$ introduces environmental stabilization, driving the effective environmental timescale slower and moving the optimal memory timescale up. Accounting for these derivatives means that 
\begin{align}
\begin{aligned}
	d_\alpha\br{1/\tildete} &= f(\ta) + \alpha \partial_{\ta}f(\ta)\partial_\alpha \ta.
\end{aligned}
\end{align}
Now, we will again make use of the assumption that $\ta$ is close $\ta^*$ such that we can make the linear approximation $f(\ta)\approx f(\ta^*) + (\ta-\ta^*)f'(\ta^*)$. Putting this in, we find
\begin{align}
\begin{aligned}
	d_\alpha\br{1/\tildete}(\ta) &= f(\ta^*) + (\ta-\ta^*)f'(\ta^*) +\\
	&\qquad \alpha \partial_{\ta}[f(\ta^*) + (\ta-\ta^*)f'(\ta^*)]\partial_\alpha \ta^*
\end{aligned}\label{si eq:grad1}
\end{align}
For a passive agent, this simplifies because $\alpha=0$. Furthermore, we know that $f'(\ta^*)=0$ because we have assumed that the agent is at optimal memory so any deviation from optimal memory must generally increase the typical distance between environmental and agent bias $(h-\he)^2$. Then,
\begin{align}
\begin{aligned}
	d_\alpha\br{1/\tildete}(\ta) &= f(\ta^*).
\end{aligned}\label{si eq:dtildete1}
\end{align}
Eq~\ref{si eq:dtildete1} is already clear from Eq~\ref{si eq:tildete1} given the assumptions we have made, but these steps take us through the general problem (when not situated exactly at optimal memory and when $\alpha\neq0$ are more complicated). In other words, decreasing $\alpha$ for the weak stabilizer will reduce the probability that the environment switches by the term in Eq~\ref{si eq:dtildete1} because $f>0$ and $f'$ --- the change in probability is not just dependent on the rate effect $f$ but also its derivative.

Under such a change, the new environmental timescale will deviate from $\te$ and so the stabilization cost can be expanded as
\begin{align}
\begin{aligned}
	G(\te,\tildete) &= \frac{1}{\te} \log \left( \frac{1/\te}{\br{1/\tildete}} \right) + \left(\frac{1}{\te}\right)\log\left( \frac{1-1/\te}{1-\br{1/\tildete}} \right)\\
		&\approx \frac{1}{2\te}\left[\br{1/\tildete}-1/\te\right]^2 +\\
		&\qquad\frac{1}{2}\left( 1-\frac{1}{\te} \right)\left[\br{1/\tildete}-1/\te\right]^2\\
		&= \frac{1}{2}\left[\br{1/\tildete}-1/\te\right]^2,
\end{aligned}
\end{align}
a cost that increases quadratically with the change in the averaged switch probability $\br{1/\tildete}$ away from $1/\te$. For a passive agent, this direction is 0 unless we allow for $\alpha$ to vary, which leads to the relation
\begin{align}
	G(\te,\tildete) &= \frac{\alpha^2}{2} f(\ta^*)^2. \label{si eq:tildete2}
\end{align}
Eq~\ref{si eq:tildete2} tells us that if we vary $\alpha$, we must pay a stabilization cost that, at least locally, grows quadratically with the strength of stabilization with zero gradient.

The simplest contribution is with respect to the change in the precision timescale $\tc$. Divergence, as derived in Materials \& Methods, is proportional to $1/\tc$. On the other hand, precision cost is $C = \log\tc$. Since optimal memory timescale does not depend on $\tc$, the change of the total divergence is
\begin{align}
	\partial_{\tc} \left[D_1/\tc + \beta \log_2\tc\right] &= -D_1/\tc^2 + \beta/\tc,
\end{align} 
where we take $D^*=D_1/\tc$ to encapsulate the terms in the divergence apart from the scaling with precision timescale. If this has a minimum at positive $\tc$, the value of $\tc$ at which the minimum is reached is $\tc^* = D_1/\beta$.

Putting all of these together, we have the terms in the gradient
\begin{align}
\begin{aligned}
	\partial_{\ta} \mathcal{D} &= 2a(\ta-\ta^*)\\
	\partial_{\alpha} \mathcal{D} &= \left[\frac{D_0}{2} (\ta^*)^{-\frac{3}{2}} + a\frac{\mu^2(2\ta^*+1)}{2\log(2)(\ta^*)^3(\ta^*+1)^3} \right] \left|\frac{\partial \ta^*}{\partial \alpha}\right| \\
	\partial_{\tc} \mathcal{D} &= \beta/\tc-D_1/\tc^2
\end{aligned}\label{si eq:gradient}
\end{align}
When the cost gradient $\partial_\alpha \mathcal{D} < 0$, a population of passive agents is driven towards niche construction and when $\partial_{\tc}\mathcal{D}<0$ towards precision reduction. Thus, the conditions that lead to reduction in agent complexity by increasing memory, enhancing stabilization, and lowering precision are captured by these gradients. 

A similar derivation can be made for the evolution of a starting population of destabilizers, or agents with $\alpha>0$, instead a pure population of passive agents. However, this requires us to deal with all the terms in Eq~\ref{si eq:grad1} and to account for a term from the gradient of stabilization cost in Eq~\ref{si eq:gradient} instead of assuming $\alpha=0$. The change in the environmental timescale is more complicated to calculate because we must then consider the way that destabilization determines the modified environmental timescale in Eq~\ref{si eq:tildete1}, but it is clear that the qualitative results will be the same because of the adaptive gain from slower environmental timescales, i.e.~decreasing $\alpha$, but the exact rate at which $\alpha$ changes will depend on the curvature of the stabilization cost.

\section{Metabolic costs of neural tissue for memory}
In the total divergence in Eq~\ref{eq:total divergence} and as discussed in Appendix~\ref{si sec:evolution}, we consider information costs separately from energetic, metabolic costs of neural tissue. An important consideration for comparing the costs directly with one another is that that the right units for comparison are not clear, an issue that we avoid by only considering the scaling exponents presented in Result 3. Furthermore, while the scaling argument makes clear that the metabolic costs will dominate at sufficiently long lifetimes, the differences in how information and energetic costs affect reproductive fitness make a direct comparison in a combined ``total divergence'' equation problematic. 

Nonetheless, if we do entertain the inclusion of metabolic costs into the total divergence, we will find that rising metabolic costs with environmental timescale will lead to a upper cutoff, i.e.~truncating memory at some point beyond which the benefits of increasing stabilization are counteracted by the monotonically increasing costs of supporting neural tissue for memory. 

To show this more formally, we redo the calculations in the previous section with an additional metabolic cost of memory,
\begin{align}
	\mathcal{D} &= \bar{D} + \mu H(\ta) + \chi G(\te,\tildete) + \beta C(\tc) + \gamma F(\ta),
\end{align}
with the new term $F(\ta)=\ta^{2\phi}$ defining the metabolic cost from Result 3. Then, we have for the shifted optimal memory
\begin{align}
\begin{aligned}
	\ta^{**} &= \ta^* +\frac{\mu}{2(\log2)a\ta^*(\ta^*+1)} - \frac{\phi (\ta^*)^{2\phi}}{a(\ta^*+1) } + \\
	&\qquad\mathcal{O}(\mu^2) + \mathcal{O}(\gamma^2) + \mathcal{O}(\mu\gamma), \label{si eq:tss}
\end{aligned}
\end{align}
obtained from $\partial_{\ta}[\bar{D} + \mu H + \gamma F]=0$ and using the approximation that both $\mu$ and $\gamma$ are small. The perturbative assumption is not necessary to take, but then there is no closed analytical solution for the shifted optimal memory $\ta^{**}$ that we can write down. Eq~\ref{si eq:tss} shows us that memory complexity, the term proportional to $\mu$, tends to drive optimal memory $\ta^{**}$ up but the metabolic cost, the term proportional to $\gamma$, tends to drive it down, the balance of which determine the exact change in optimal memory.

Taking the approximation in Eq~\ref{si eq:tss} the shifted optimal divergence, denoted by an apostrophe, becomes
\begin{align}
\begin{aligned}
	\bar{D}'(\ta^{**}) &= D^* + a\frac{\mu^2}{4(\log2)^2(\ta^*)^2(\ta^*+1)^2} + \\
		& \quad\frac{\gamma^2\phi^2 (\ta^*)^{4\phi}}{a^2(\ta^*+1)^2} - \frac{a\phi\mu\gamma (\ta^*)^{2\phi}}{2(\log2)a\ta^*(\ta^*+1)^2} + \\
		&\quad\mathcal{O}(\mu^3) + \mathcal{O}(\gamma^3) + \mathcal{O}(\mu\gamma^2) + \mathcal{O}(\mu^2\gamma).
\end{aligned}
\end{align}
Again, perturbations about the local optimum lead to
\begin{align}
\begin{aligned}
	\bar{D}'(\ta) &\approx D^* + a\frac{\mu^2}{4(\log2)^2(\ta^*)^2(\ta^*+1)^2} + \\
	& \quad\frac{\phi^2 (\ta^*)^{4\phi}}{a^2(\ta^*+1)^2} - \frac{a\phi\mu (\ta^*)^{2\phi}}{2(\log2)a\ta^*(\ta^*+1)^2} + \\
	& \quad b(\ta-\ta^{**})^2\label{si eq:shifted div}
\end{aligned}
\end{align} 
for some positive constant $b$, which implicitly depends on the complexity cost. Assuming that the population is at optimal memory, we obtain for the partial derivative with respect to $\alpha$
\begin{align}
\begin{aligned}
	\partial_\alpha \bar{D}' &= -\left[\frac{D_0}{2} (\ta^*)^{-3/2} + a\frac{\mu^2(2\ta^*+1)}{2(\log2)(\ta^*)^3(\ta^*+1)^3} + \right. \\
	& \frac{2(\log2)\phi^2a^{-2}(\ta^*)^{4\phi}+\mu\phi(\ta^*)^{2\phi-1}}{(\log2)(\ta^*+1)^3} - \\
	&\left. \frac{4(\log2)\phi^3(\ta^*)^{4\phi-1} + 2^{-1}\mu\phi(2\phi-1)(\ta^*)^{2\phi-2}}{(\log2)(\ta^*+1)^2} \right] \frac{\partial \ta^*}{\partial \alpha}.\\
\end{aligned}\label{si eq:D star2}
\end{align}
Unlike the previous outcome in Eq~\ref{si eq:D star}, it is not necessarily the case that a stronger stabilizer will decrease divergence because sufficiently large metabolic costs will counteract the adaptive benefits of a slower environment.

\begin{table*}\centering
\caption{Variables used in main text organized by section in which they are first introduced or used.}\label{si tab:notation}
\begin{tabular}{rl}
	\bf Model structure \& assumptions \\
	Parameter & Description \\
\hline
	$A_t$ &	discrete agent state at time $t$, e.g.~$\{-1,1\}$\\
	$E_t$ &	discrete environmental state at time $t$, e.g.~$\{-1,1\}$ \\
	$h_0$ &	parameter for strength of environmental bias \\
	$h$ &	agent bias \\
	$\hat h$ &	agent's estimate of environmental bias	\\
	$h_E$ &		environmental bias	\\
	$p$ &	agent's probability distribution over possible states of $A_t$ after time integration	\\
	$\hat p$ &	agent's estimate of environment probability distribution at time $t$ based on present samples	\\
	$p_E$ &	environmental probability distribution over possible states of $E_t$ \\
	$q$ &	probability of change in environmental bias at a single time step	\\
	$s$ &	state of environment taking values of $-1$ or $1$	\\
	$t$ &	time	\\
	$v$ &	construction rate curvature	\\
	$\alpha$ &	construction rate weight, $\alpha<0$ for stabilizers and $\alpha>0$ for destabilizers	\\
	$\beta$ &	learning weight in Eq 4; coefficient of precision cost in Eq 18	\\
	$\epsilon_{\tau_c}$ &	perceptual error	\\
	$\eta_{\tau_{\rm c}}$ &	estimated bias error	\\
	$\tau_{\rm c}$ &	sampling duration, inverse precision	\\
	$\tau_{E}$ &	environment duration	\\
	$\tau_{\rm f}$ &	niche construction duration \\
	$\tau_{\rm m}$ &	agent memory duration	\\
	\\
	\bf Result 1 \\
	Parameter & Description \\
\hline
	$\bar D$ &	time-averaged Kullback-Leibler (KL) divergence	\\
	$\bar D^*$ &	time-averaged KL divergence at optimal memory duration	\\
	$D_{\rm KL}$ &	KL divergence	\\
	$\tau_{\rm m}^*$ &	optimal memory duration	\\
	\\
	\bf Result 3 \\
	Parameter & Description \\
\hline
	$B$ & 	metabolic rate	\\
	$M_{\rm br}$ &	brain mass	\\
	$N$ &	number of episodes of environmental change	\\
	$T$ &	lifespan of organism	\\
	$y$ &	exponent relating environment duration and organism lifetime	\\
	$\phi$ &	exponent relating metabolic rate and memory duration, $\phi=a/4b$ for energetic exponents $a$ and $b$	\\
	\\
	\bf Result 4 \\
	Parameter & Description \\
\hline
	$C$ &	sensory precision cost	\\
	$\mathcal{D}$ &	total divergence	\\
	$G$	&	stabilization cost	\\
	$H$ &	complexity of memory cost	\\
	$\beta$	&	coefficient for sensory cost \\
	$\mu$ &	coefficient for memory complexity cost	\\
	$\tilde{\tau}_{E}$ &	modified environment duration	\\
	$\chi$ &	coefficient for stabilization cost
\end{tabular}
\end{table*}

\clearpage
\bibliography{refs,jrefs}

\end{document}